\documentclass[
reprint,
amsmath,amssymb,
pra,
aps,
]{revtex4-2}

\usepackage{natbib}
\usepackage{graphicx}
\usepackage{dcolumn}
\usepackage{bm}
\usepackage{physics}
\usepackage{nicefrac}
\usepackage[caption=false]{subfig}
\usepackage{xcolor}
\usepackage{lineno}

\begin{document}

\preprint{APS/123-QED}

\title{Improved autonomous error correction using variable dissipation in small logical qubit architectures}

\author{David Rodr\'iguez P\'erez}
 \email{drodriguezperez@mines.edu}
\author{Eliot Kapit}%
\affiliation{%
Department of Physics, Colorado School of Mines, Golden, CO 80401
}%
\affiliation{
Department of Physics and Engineering Physics, Tulane University, New Orleans, LA 70118
}



\begin{abstract}
Coherence times for superconducting qubits have greatly improved over time. Moreover, small logical qubit architectures using engineered dissipation have shown great promise for further improvements in the coherence of a logical qubit manifold comprised of few physical qubits. Nevertheless, optimal working parameters for small logical qubits are generally not well understood. This work presents several approaches to finding preferential parameter configurations by looking at three different cases of increasing complexity. We begin by looking at state stabilization of a single qubit using dissipation via coupling to a lossy object. We look at the limiting factors in this approach to error correction, and how we address those by numerically optimizing the parametric coupling strength with the lossy object having an effective time-varying dissipation rate---we call this a pulse-reset cycle. We then translate this approach to more efficient state stabilization to an abstracted three-qubit flip code, and end by looking at the Very Small Logical Qubit (VSLQ). By using these techniques, we can further increase logical state lifetimes for different architectures. We show significant advantages in using a pulse-reset cycle over numerically optimized, fixed parameter spaces.
\end{abstract}

\maketitle


\section{\label{sec:level1}Introduction}

Error correction via the encoding of logical qubits using multiple physical qubits is a very promising route towards fault-tolerant quantum computation \cite{Fowler2012}. While topological stabilizer codes like the surface code propose a blueprint to implement quantum error correction, they also require many physical qubits to encode a single logical qubit. For a distance-3 surface code, one logical qubit is encoded using 17 physical qubits \cite{Fowler2012}. Concurrently, research using engineered dissipation for stabilizing quantum states has been growing \cite{Kapit2017, Ma2019, Gertler2020, Kristensen2019}. Some prominent examples are cat codes \cite{Leghtas2013,Mirrahimi2014,Sun2014,Leghtas2015,Albert2016,Ofek2016,Cohen2017,Mundhada2017,Michael2016,Wang2016,Heeres2017,Puri2017}, having already exceeded break-even \cite{Ofek2016}, and the Very Small Logical Qubit (VSLQ) \cite{Kapit2016} where a logical qubit is encoded using only four physical qubits, coupling two high-coherence qubits each with two lossy qubits or resonators.

By modulating the coupling between two quantum devices, red-sideband photon exchange interactions $( a_1a_2^{\dagger} + a_1^{\dagger}a_2)$ \cite{Beaudoin2012, Strand2013,Roth2017} and blue-sideband photon squeezing $( a_1a_2 + a_1^{\dagger}a_2^{\dagger})$ \cite{Roth2017,Wallraff2007,Leek2009,Novikov2016} can be achieved for autonomous state stabilization \cite{Lu2017, Huang2018}. The goal of this paper is to better characterize and understand the limiting error channels using the blue-sideband two photon creation and annihilation coupling used in our autonomous error correction schemes, and eliminate them with a pulse-reset technique. This is done with a numerically optimized, time-parameterized coupling strength with a time-varying lossy object. In the next section, we talk about how the use of quantum noise as engineered dissipation aids in the stabilization of a quantum state, which is the fundamental mechanism by which these small logical devices achieve autonomous error correction. We then discuss the limiting error syndrome induced by the error correction mechanism itself, whereby the coupling strength between the high-coherence and the lossy qubits may induce off-resonant transitions into unwanted leakage states. To minimize this effect, we introduce techniques borrowed from gate optimization by numerically optimizing a parametric, time-varying coupling strength, and qubit reset protocols that allow the lossy qubit enough time to reset as shown in Fig.~\ref{fig:pulseReset} \cite{Magnard2018, Reed2010,Valenzuela2006,Geerlings2013}.

We begin by looking at the simplest, idealized scenario of a single high-coherence, three-level qubit device with an unwanted leakage state, coupled to a single lossy qubit or oscillator to understand the stabilization mechanism and the limiting error. The numerical optimization of the time-parameterized coupling and the qubit reset protocols show a very clear advantage over a standard, fixed-coupling parameter space for single-qubit device state stabilization. This is done without assuming any specific architecture, the only criteria being having the ability to couple two qubit devices with strong coupling manipulation, and efficiently inducing qubit reset. To show the practicality of these techniques, we generalize its use on a three-qubit flip code, in which we look at how to optimize autonomous correction of a qubit-flip error syndrome. Having shown the improvement from using these techniques, we proceed to the more complex VSLQ, where applying these methods gives us a better performance over numerically optimizing the individual parameters in the system Hamiltonian.

\section{\label{sec:singleQubit} Single-qubit stabilization}

\subsection{\label{sec:sq-system} System}

\begin{figure*}
	\includegraphics[width = \textwidth]{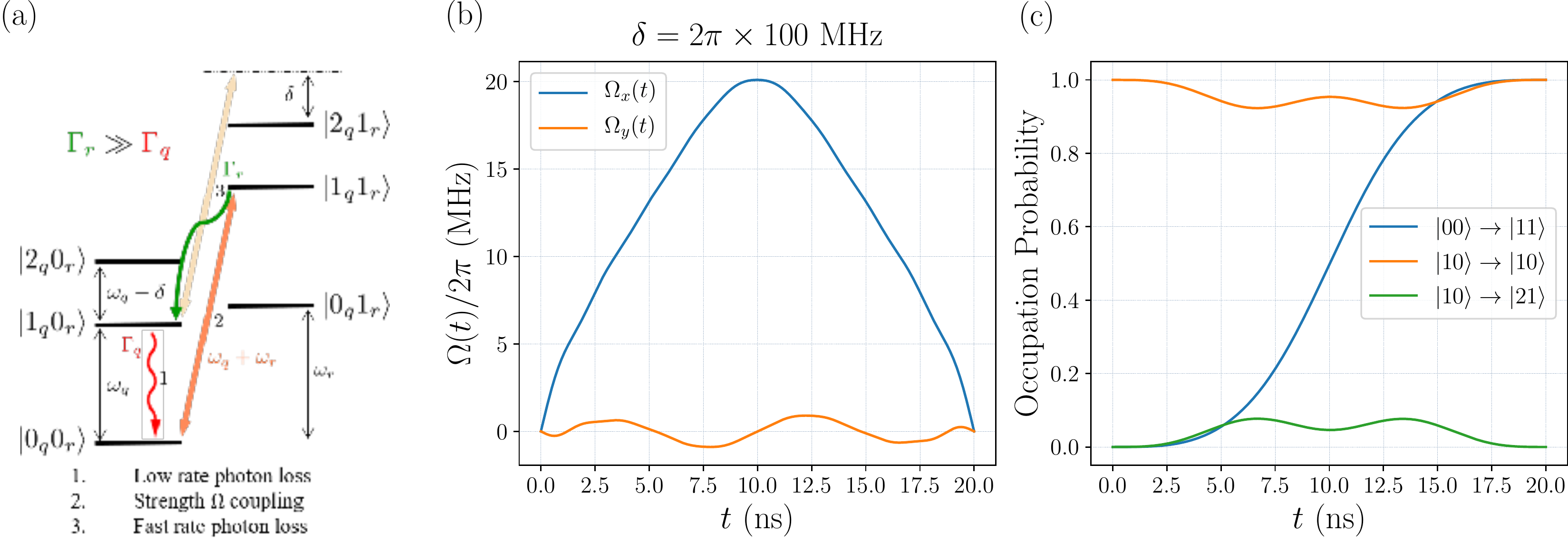}
	\caption{\label{fig:singleQubit} A high-coherence qubit device coupled to a lossy qubit or resonator with a coupling strength $\Omega$. (a) An energy level diagram of the system. We only consider the ground state and first excited state of the resonator due to its lossy nature, making occupations of any higher states extremely unlikely. The protocol follows a photon loss in the primary qubit device (1), followed by the coupled excitation of both qubit device and resonator at strength $\Omega$ (2), and finally the relaxation of the resonator (3). (b) Optimized pulse shape for the operation $\ket{00} \rightarrow \ket{11}$ from a gradient ascent optimization. We let $N=20$ in Eq.~(\ref{eq:omega}), and initialize $c_1^x= 2\pi\times 20\ \text{MHz}$, $c_{n \neq 1}^x = 0$, and all $c_n^y = 0$, letting them vary by $\epsilon$ until we achieve a target state fidelity of 0.9989. All photon loss is turned off in this optimization, with the goal of trying to minimize errors induced by this mechanism itself. (c) We track the fidelities of the transitions $\ket{00}\rightarrow\ket{11}$ (blue), $\ket{10}\rightarrow\ket{10}$ (orange), and $\ket{10}\rightarrow\ket{21}$ (green) from the qubit-resonator coupling. The goal is to excite both qubit device and resonator in the event of a photon loss, leaving the target state unchanged while minimizing off-resonant transitions into the leakage state.}
\end{figure*}

For a more thorough explanation on using engineered dissipation, we direct the reader to a full review on the subject \cite{Kapit2017}. We look at a very simple, idealized example similar to \cite{Lu2017} as a warm-up show case for the more interesting and complex applications of error correction in the next sections. Consider a single high-coherence qubit device coupled to a single lossy qubit or resonator. Here, we consider the high-coherence qubit device to be a three-level system with a nonlinearity $\delta$, and the lossy resonator to be a two-level system, as shown in Fig.~\ref{fig:singleQubit}(a). The lossy nature allows us to truncate the resonator to its first two energy states, since occupation of any higher states are extremely unlikely and do not contribute any meaningful dynamics. This also has the added benefit of faster simulations due to the smaller Hilbert space. For this example, our only focus is to stabilize a single, excited state for the high-coherence qubit device in the computational basis. However, engineered dissipation can be used to stabilize an arbitrary state along any axis on the Bloch sphere, as described in \cite{Huang2018}.

\begin{figure}
	\includegraphics[width = 0.485\textwidth]{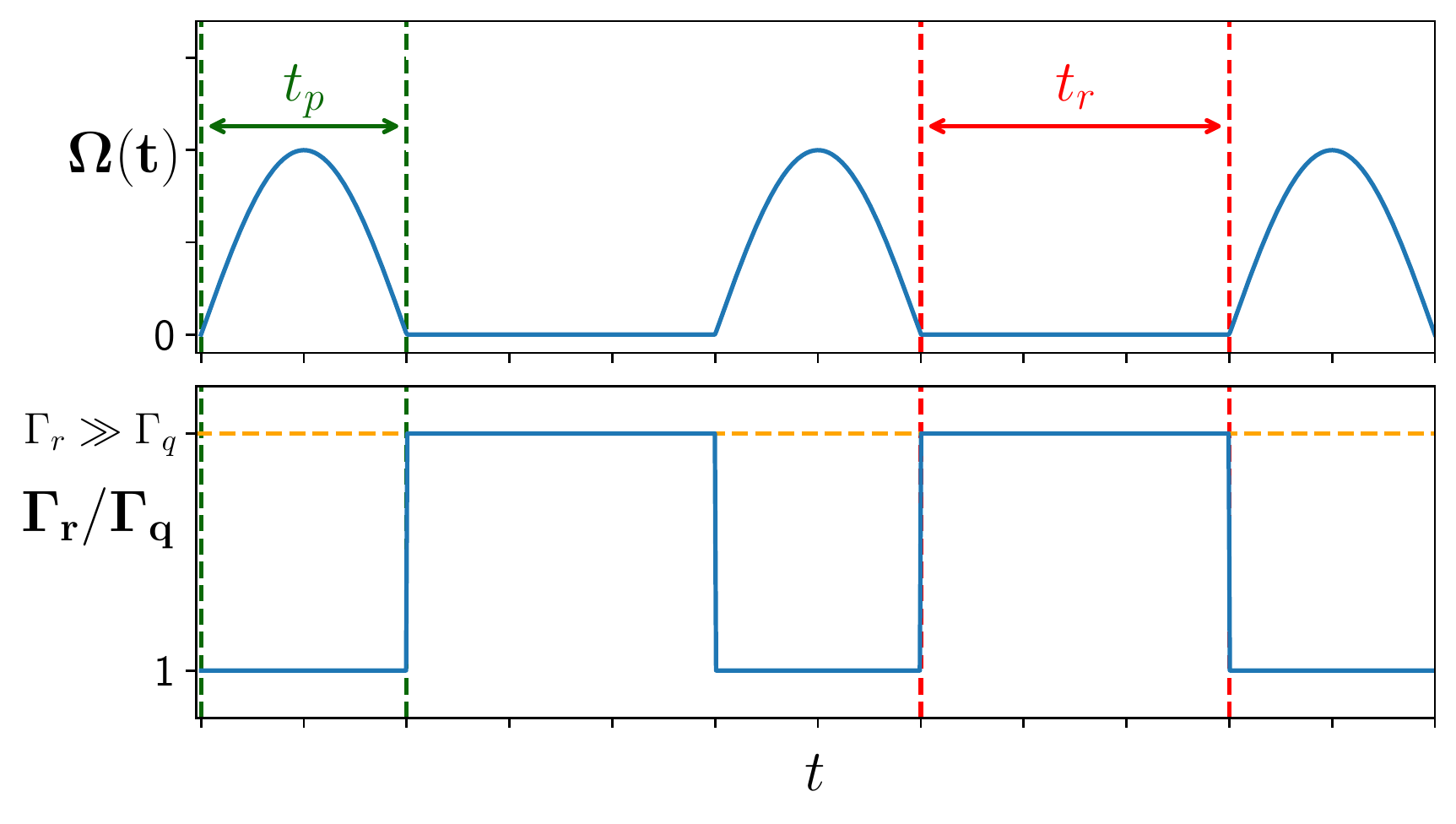}
	\caption{\label{fig:pulseReset} Scheme for the pulse-reset cycles. We denote the coupling duration as $t_p$ (green), in which the qubit device and resonator are coupled using an optimized pulse shape. The reset cycle, $t_r$ (red), is determined using a simple scan over different values to determine what gives the lowest residual error rate for each different $T_1$. During $t_p$ we set $\Gamma_r = \Gamma_q$ and $\Omega = \Omega_{opt}(t)$, while during $t_r$ we have $\Gamma_r \gg \Gamma_q$ and $\Omega = 0$. It is assumed that qubit reset protocols, such as the one described in \cite{Magnard2018}, can be performed efficiently, giving us an effective $\Gamma_r \gg \Gamma_q$.}
\end{figure}

We will refer to the high-coherence qubit device as simply the primary qubit, and the ancilla as just a resonator. The system is described by the Hamiltonian
\begin{equation}\label{eq:singleQubitOGHam}
	\begin{split}
		H = \frac{\omega_q}{2}\sigma^z_q &- \delta P_q^2 + \omega_r a_r^{\dagger}a_r \\
		&+ 2\Omega\cos{2\omega_0t} \left( a_q + a_q^{\dagger} \right)\left( a_r + a_r^{\dagger} \right).		
	\end{split}
\end{equation}
Moving to the rotating frame under the unitary transformation $U(t) = e^{-i(\omega_rn_r + \frac{1}{2}\omega_q\sigma^z_q)t}$, and discarding counter rotating terms as well as choosing an appropriate modulation of the coupling strength, we can write the rotating wave Hamiltonian of this system as

\begin{equation}\label{eq:singleQubitHam}
	H = -\delta P_q^2 + \Omega \left(a_qa_r + a_q^{\dagger}a_r^{\dagger}\right).
\end{equation}
Here, $P_q^2$ is the projector operator $\ketbra{2_q}{2_q}$ onto the leakage state of the qubit, and $\Omega$ is the strength of the blue-sideband coupling between the qubit and resonator.

The dynamics of the autonomous error correction protocol for this example is as follows. In the event of a photon loss in the qubit, $\ket{1_q0_r} \rightarrow \ket{0_q0_r}$, the coupled drive will excite both the qubit and resonator into their first excited states, $\ket{0_q0_r} \rightarrow \ket{1_q1_r}$, after which the lossy resonator will bring the system back to the initial desired state, $\ket{1_q1_r} \rightarrow \ket{1_q0_r}$. The important conditions here are: 1. $\Gamma_q \ll \Gamma_r$, meaning the lossy resonator can quickly decay  back to its ground state, bringing the entire system back to the initial state $\ket{1_q0_r}$, and 2. $\delta \gg \Omega$, this reduces unwanted off-resonant transitions into the primary qubit leakage state induced by the coupling with the resonator. While this scheme proves very efficient in achieving autonomous error correction, off-resonant transitions induced by the error correction mechanism itself present a limiting factor \cite{Ma2017}.

To that end, we use techniques in pulse engineering \cite{Khaneja2005, Motzoi2009, Mottonen2006, Steffen, Safaei2009, Pravia2003} to eliminate leakage during the qubit-resonator coupling, as well as a reset cycle which allows the resonator time to reset back to its ground state. After the pulsed operation $\ket{0_q0_r} \rightarrow \ket{1_q1_r}$, there is a non-zero probability that the primary qubit may undergo another photon loss before the resonator has decayed back to its ground state, $\ket{1_q1_r} \rightarrow \ket{0_q1_r}$. In this event, the coupling terms perform the transition $\ket{0_q1_r} \rightarrow \ket{1_q2_r}$, which is far off-resonant and less likely to accomplish the goal of taking the primary qubit to its first excited state. Thus the reset cycle ensures the resonator will be usable for correcting the qubit to $\ket{1_q}$. These two steps, laid out in in Fig.~\ref{fig:pulseReset} are imperative to our approach of eliminating the limiting error channels of the standard, continuous coupling schemes.

\subsection{\label{sec:sq-results} Results}

We borrow gate optimization techniques from \cite{Motzoi2009} by pulse shaping an analogous two quadrature coupling strength, letting $\Omega \rightarrow \Omega_x(t), \Omega_y(t)$ [Fig.~\ref{fig:singleQubit}(b)], where
\begin{equation}\label{eq:omega}
	\Omega_{x,y}\left(t\right) = \displaystyle\sum_{n=1}^{N} c^{x,y}_n\sin\left(n\pi t/t_p\right),
\end{equation}
with $\Omega_x(t)$ corresponding to the coupling term in Eq.~(\ref{eq:singleQubitHam}), and another pulse component $\Omega_y(t)$ to eliminate leakage influenced by the DRAG protocol, giving us the effective Hamiltonian
\begin{equation}\label{eq:singleQubitHamEff}
	\begin{split}
		H_{\textrm{eff}} = -\delta P_q^2 &+ \Omega_x(t)\left( a_qa_r + a_q^{\dagger}a_r^{\dagger} \right) \\ &+ \Omega_y(t)i\left( a_q^{\dagger}a_r^{\dagger} - a_qa_r \right).
	\end{split}
\end{equation}
Using Eq.~(\ref{eq:omega}) allows us to set the conditions $\Omega_{x,y}(0) = \Omega_{x,y}(t_p) = 0$, while allowing us to use a gradient ascent numerical optimization over the $c^{x,y}_n$ coefficients for pulse shaping the $\Omega_{x,y}(t)$ terms. By letting $c_1^x$ be some initial amplitude, and all other $c_{n\neq 1}^x,c_n^y = 0$, we find the optimized fidelity of the operation $\ket{00} \rightarrow \ket{11}$ so as to maximize the target state transition, while minimizing off-resonant transitions into the leakage state induced by the coupling itself. This is evident in Fig.~\ref{fig:singleQubit}(c), where we can see the dips in the fidelity tracking for $\ket{10}\rightarrow\ket{10}$ and $\ket{10}\rightarrow\ket{21}$ (orange and green, respectively).

In addition to optimizing the pulse shape of the coupling strength, we use alternating pulse-reset cycles in which the blue-sideband coupling strength is determined by the optimized pulse shapes during the coupling cycles. During the reset cycles, the coupling strength is completely turned off while increasing the resonator decay rate so as to allow the resonator to reset back to its ground state. This effective, induced resonator reset can be achieved with techniques described in \cite{Magnard2018}. For simplicity of simulation, we assume that this reset protocol can be achieved with high fidelity and treat the decay rate as time-varying, which is high during the reset cycle and low during a coupling cycle (Fig.~\ref{fig:pulseReset}). We simulate these dynamics, as well as the following, more complex examples, with a continuous time evolution of Schr\"odinger equations for the pulse optimizations, and Lindblad master equations for the full pulse-reset cycle evolutions. For this single qubit stabilization example, we use the Hamiltonian Eq.~(\ref{eq:singleQubitHamEff}) over a $\left(3 \times 2\right)$-dimensional Hilbert space [Fig.~\ref{fig:singleQubit}(a)], and collapse operators $a_{q,r}$ with rates $\Gamma_{q,r}$.

\begin{figure}
	\includegraphics[width = 0.48\textwidth]{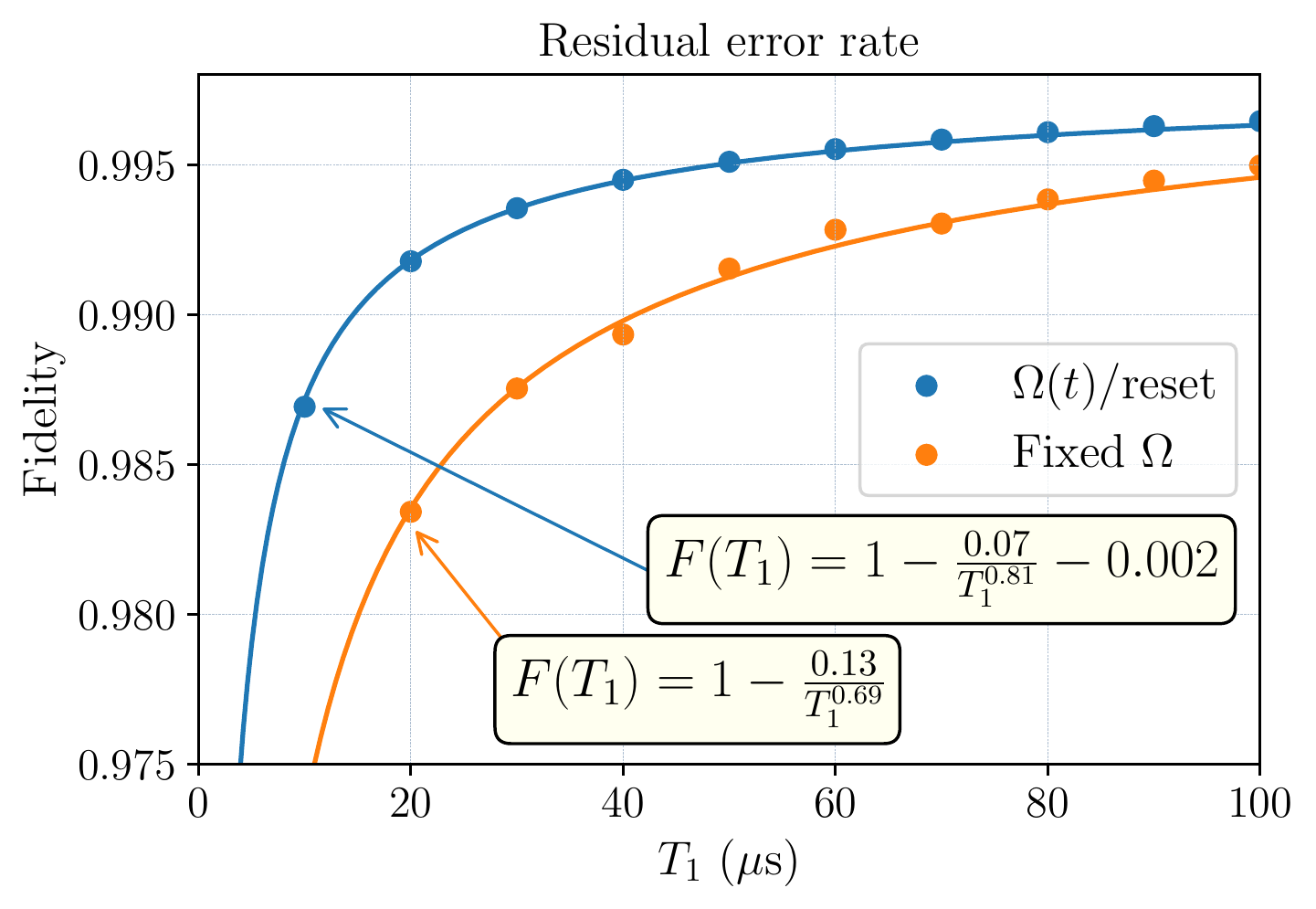}
	\caption{\label{fig:residualError} Residual error rate scaling with the fidelity calculated at the end of one pulse-reset cycle. We use individually preferred $t_r$ for each $T_1$, and the same optimized $\Omega_{opt}(t)$ (Fig.~\ref{fig:singleQubit}(b)). We report a residual error scaling $\propto T_1^{-0.81}$, compared to $T_1^{-0.69}$ using constant coupling.}
\end{figure}

We report a target state fidelity of 0.9989 [Fig.~\ref{fig:singleQubit}(c)] for the operation $\ket{00} \rightarrow \ket{11}$, using the resulting pulse shape in Fig.~\ref{fig:singleQubit}(b). However, it is important to note that experimental results in \cite{Lu2017} achieve stabilization of the qubit excited and ground states with $>90\%$ and $>99\%$ purity, respectively. Meanwhile, stabilization of an arbitrary state along the Bloch sphere achieves a worst case average fidelity exceeding 80\%. This apparent discrepancy is because the results presented in Fig.~\ref{fig:singleQubit}(b) and Fig.~\ref{fig:singleQubit}(c) are obtained without accounting for decoherence. This was done because the point is to reduce the limiting error induced from off-resonant transitions during the qubit-resonator coupling \cite{Ma2017}, and so the $\Omega_{x,y}(t)$ optimization is done without considering photon losses. The results of the residual error rate scaling using constant coupling strength versus optimized time-varying coupling strength and reset cycles are summarized in Fig.~\ref{fig:residualError}. While here we do consider photon loss for the full pulse-reset evolution, we still do not consider phase noise in an effort to keep this a simplified example of the use of this technique. Further considerations for a more complete picture of single qubit stabilization, as well as a discussion on the broad range of factors to improve on experimental results, would introduce increasing complexity and would be a deviation from the main purpose of this paper.

For our comparisons, we let the fidelity $F \rightarrow F(T_1,\Omega, \Gamma_r)$, where for each different $T_1$, $F$ is gradient-ascent optimized to find the best choice of $\Omega$ and $\Gamma_r$. While we do see a much improved scaling $\propto T_1^{-0.81}$, it is still short of the theoretical optimal $\propto T_1^{-1}$ \cite{Ma2017}. However, given experimental limitations, this improvement is highly desirable over the asymptotically fixed coupling strength error scaling $\propto T_1^{-\nicefrac{1}{2}}$, which is especially evident for short $T_1$. We note that the scaling for the data using fixed coupling strength (orange curve) does not agree with the expected results from \cite{Ma2017}, which may be a result of the theoretical scaling being asymptotic, and not visible in the range of $T_1$ values explored. Likewise, while we expect the pulse-reset data scaling to approach unity (blue curve), the presence of a $0.002$ offset may suggest the technique itself is a limiting factor, or just another problem with the range of explored $T_1$ values. Nevertheless, we can still very clearly see a noticeable advantage in using pulse-resets over fixed coupling for short coherence times $T_1$, providing a meaningful showcase for the use of this technique.

\begin{figure}
	\includegraphics[width = 0.48\textwidth]{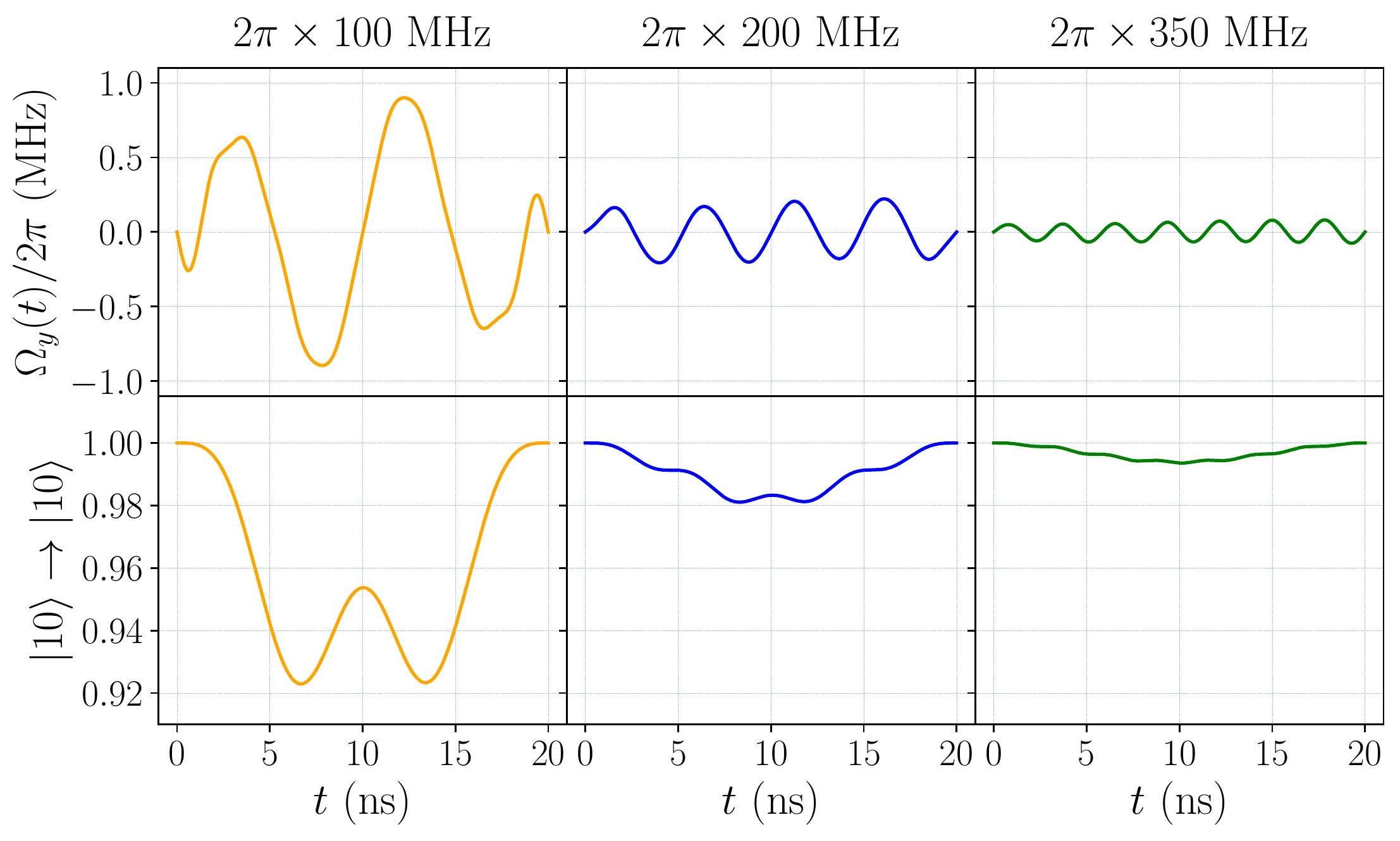}
	\caption{\label{fig:yterms} The effects of the $\Omega_y(t)$ terms on off-resonant, blue-sideband transitions for $\delta=2\pi\times\{100,200,350\}\ \textrm{MHz}$. There is a clear correlation between the effective oscillating counterterm and the size of $\delta$. This is shown by tracking the occupation of the target state $\ket{1_q0_r}$ (bottom plots), as well as the leakage state $\ket{2_q1_r}$ in Fig.~\ref{fig:singleQubit}(c), having initialized in the target state $\ket{1_q0_r}$.}
\end{figure}

We draw attention to the shape of the $\Omega_y(t)$ pulse as a result of the gradient ascent search. From Fig.~\ref{fig:yterms}, we see that the oscillatory behavior changes for different nonlinearity $\delta$. Indeed, from Fig.~\ref{fig:singleQubit}(a) it is clear that having a larger nonlinearity reduces the probability for off-resonant transitions into the leakage state induced by the coupling strength. For $\delta=2\pi \times \{100,200,350\}\ \textrm{MHz}$, we see that the frequency of this $\Omega_y(t)$ counterterm strength is about 100, 200 and 350 MHz, respectively. This opens up the possibility of a more general use of oscillatory counterterms in driven fields for state stabilization, and further research will be required to understand this phenomenon.

\section{\label{sec:threeQubit} Three-qubit flip code}

\subsection{\label{sec:tq-system} System}

The single-qubit case we just discussed is a very simple demonstration of the potential in using this time-parameterized coupling technique for state stabilization. We now present the use of this technique in a more complex, yet still abstracted, system---the three-qubit flip code \cite{Reed2012}. We consider a case proposed in \cite{Cohen2014} and \cite{Kapit2014} where we have 6 total qubits, three high-coherence qubits each coupled to three lossy qubits, with the primary qubits coupled to each other, and the lossy resonators decoupled from each other but coupled to a primary qubit each. Following the bit-flip code laid out in \cite{Kapit2014}, by appropriately tuning flux-biased Josephson junction couplings between the primary qubits, and coupling the lossy qubits with a weak capacitive interaction with their respective primary qubits, we can achieve a rotating-frame Hamiltonian of the Jaynes-Cummings form separating the system Hamiltonian into three parts for the high-coherence qubits (denoted as the primary qubits Hamiltonian $H_P$ as in \cite{Kapit2017}), the lossy resonators $H_R$, and the interaction Hamiltonian for the coupling between the lossy and high-coherence qubits $H_{PR}$. We have
\begin{equation}\label{eq:threeQubit}
	\begin{split}
		H &= H_P + H_R + H_{PR},\\
		H_P &= -J\left( \sigma_{1P}^z\sigma_{2P}^z + \sigma_{2P}^z\sigma_{3P}^z + \sigma_{1P}^z\sigma_{3P}^z \right),\\
		H_R &= -2J\displaystyle\sum_{i=1}^3 \sigma_{iR}^z,\ H_{PR} = \Omega \displaystyle \sum_{i=1}^3\sigma_{iP}^x\left(\sigma_{iR}^x + \sigma_{iR}^y\right),
	\end{split}
\end{equation}
with $J$ being the energy scale.

While the idea is exactly the same as the single-qubit stabilization, there are some things to keep in mind that change the outcome of our goal. For this system, we are not trying to stabilize a single state but rather a logical manifold. Namely, we are protecting either logical state $\ket{0_L}$ or $\ket{1_L}$, which are defined by the majority vote of the three primary qubits,
\begin{equation}
	\begin{split}
		\ket{000},&\ket{100},\ket{010},\ket{001} \rightarrow \ket{0_L} \\ 
		\ket{111},&\ket{011},\ket{101},\ket{110} \rightarrow \ket{1_L}.
	\end{split}
\end{equation}
For either logical state, two flip errors on physical qubits is a logical flip error.

Again, just like we got to Eq.~(\ref{eq:singleQubitHamEff}), we look at replacing $\Omega$ in Eq.~(\ref{eq:threeQubit}) with a two quadrature parametric coupling $\Omega(t)$ from Eq.~(\ref{eq:omega}), and optimizing it with a gradient ascent. This system is significantly different, however, in that it is more abstracted. Rather than trying to correct photon losses in the primary qubits, we are addressing bit-flip errors. With this error syndrome, we let the Lindblad operators and error rates for the system be $\sigma^x,\ \Gamma_P$ for the primary qubits and $\sigma^{-},\ \Gamma_R$ for the lossy qubits. Additionally, because we constrain the system to bit-flip operations instead of ladder operators as collapse operators for the primary qubits, and to maintain a smaller Hilbert space, we only consider the computational space and ignore higher energy leakage states ($\ket{2}, \ket{3} ...$) for all 6 qubits. Whereas for the single-qubit system we make use of optimized coupling strength pulse shapes to perform a target operation with high fidelity while eliminating off-resonant transitions into these higher energy leakage states, here we are trying to eliminate off-resonant qubit flips induced by the coupling mechanism itself, for example, $ \ket{000}_P \otimes \ket{000}_R \rightarrow \ket{100}_P \otimes \ket{100}_R $ in attempting to protect $\ket{0_L}$. And so, this is a very useful example in demonstrating autonomous, high fidelity, time-parameterized coupling strength for error correction while eliminating unwanted off-resonant operations.

The goal of applying this lossy coupling technique is, therefore, to protect the logical states by autonomously correcting flip errors corresponding to the appropriate parent logical state. So we can't simply just flip any $\ket{0}$ states to $\ket{1}$ like we did for the single-qubit stabilization, since our target operation will depend on which logical state we are trying to protect. The target operations for $\Omega_x(t)$ and $\Omega_y(t)$ would be such that the majority is respected. For example,
\begin{equation}\label{eq:threeQubit-targets}
	\begin{split}
		\ket{100}_P \otimes \ket{000}_R &\rightarrow \ket{000}_P \otimes \ket{100}_R\\ 
		\ket{101}_P \otimes \ket{000}_R &\rightarrow \ket{111}_P \otimes \ket{010}_R,
	\end{split}
\end{equation}
where our time-dependent interaction Hamiltonian would be
\begin{equation}
	H_{PR} = \displaystyle\sum_{i=1}^3 \sigma_{iP}^x\left( \Omega_x(t)\sigma_{iR}^x + \Omega_y(t)\sigma_{iR}^y \right).
\end{equation}
Note from Eq.~(\ref{eq:threeQubit-targets}) that these operations are achieved by the red- and blue-sideband couplings described in \cite{Huang2018,Kapit2015}.

We emphasize that like the single-qubit stabilization example, this system is still idealized and abstracted. A more thorough characterization that includes leakage states would be needed for a more realistic demonstration of using a pulse-reset evolution with a qubit-flip error syndrome. However, as we saw for the single-qubit case, it is safe to assume that with a large enough nonlinearity, and with the oscillating $\Omega_y(t)$ counterterm having numerically optimized a time-parameterized coupling strength, we can reasonably exclude leakage transitions from our simulations.

\subsection{\label{sec:tq-results} Results}

\begin{figure}
	\includegraphics[width = 0.485\textwidth]{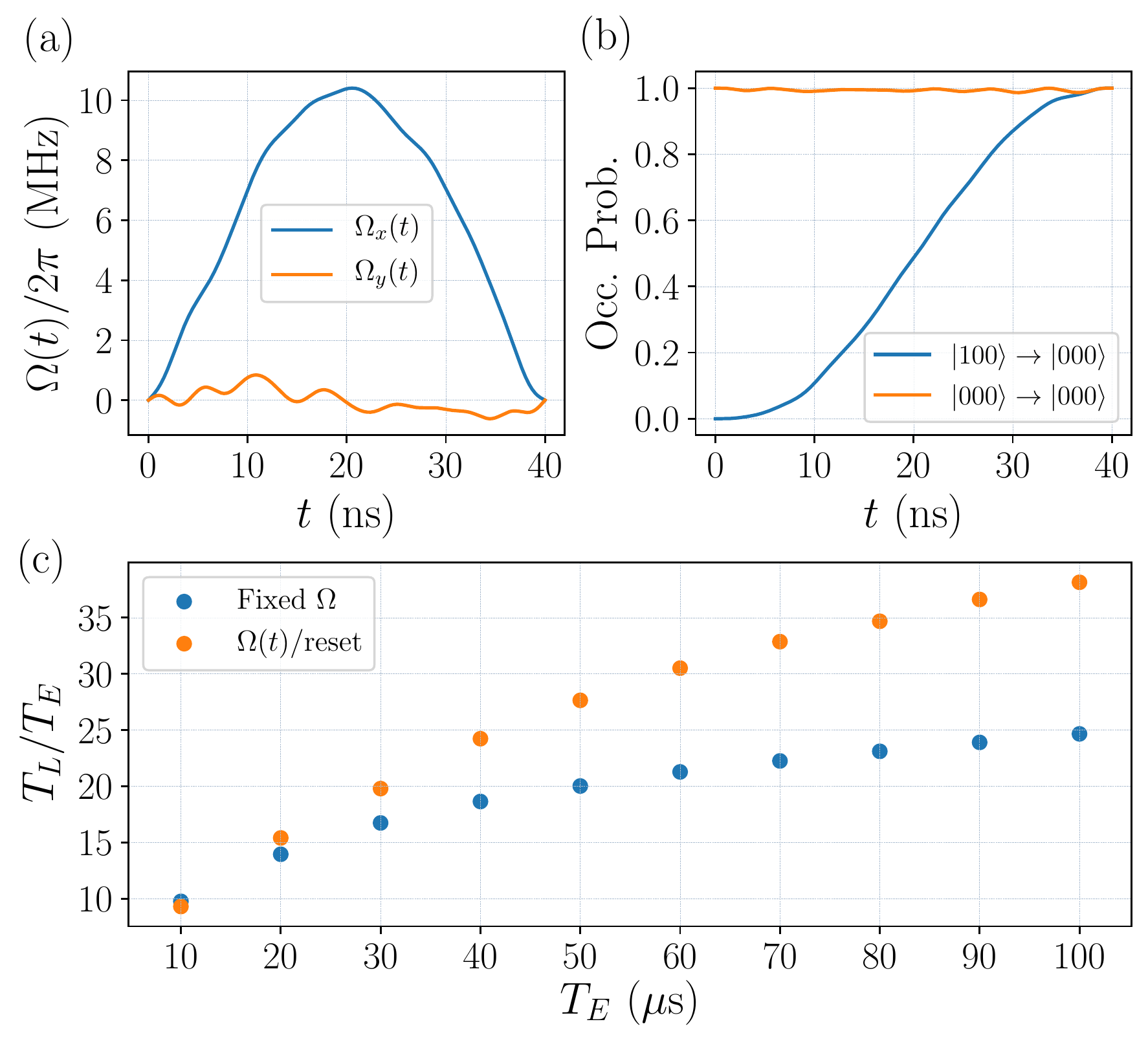}
	\caption{\label{fig:threeQubit-summary} Results for the three qubit code using $J=2\pi\times20$ MHz and $\Gamma_R = 30\ \mu\textrm{s}^{-1}$. (a) The optimized pulse shape for the target operations in Eq.~(\ref{eq:threeQubit-targets}), as well as (b) tracking the occupation probability for $\ket{000}_P$, where a final operation fidelity of 0.99999635 is achieved. Note that the states in the legend are only the primary qubits subspace and that the operation evolution is without decoherence. (c) Improvement factor $T_L/T_E$ for the states $\ket{000}_P$ and $\ket{111}_P$, with increasing error times $T_E$. The state evolutions for these last  results do include decoherence. Results in (a) and (b) were obtained with no bit-flip errors, while results in (c) do include bit=flip errors.}
\end{figure}

We summarize the results for this system in Fig.~\ref{fig:threeQubit-summary}. The $\Omega_y(t)$ pulse shape in Fig.~\ref{fig:threeQubit-summary}(a) does not have the same oscillatory behavior as we saw for the single-qubit case, which is consistent with our exclusion of a leakage state and thus no nonlinearity to correlate to. Instead, we see a pulse shape that more so resembles the time derivative of the $\Omega_x(t)$ term, with the exception of the condition $\Omega_y(0) = \Omega_y(t_p) = 0$. This shows us another example of the benefit of a second pulse quadrature in the coupling strength in an effort to eliminate unwanted, off-resonant flip operations. While we still see a very small probability of unwanted transitions out of the target logical state $\ket{000}$ during the pulse evolution, seen in Fig.~\ref{fig:threeQubit-summary}(b), we report a very good target operation Eq.~(\ref{eq:threeQubit-targets}) fidelity of $1-F < 10^{-6}$. We compare the logical lifetimes $T_L$ to single qubit error times $T_E$, as opposed to $T_1$, where $T_E = 1/\Gamma_P$ is the single qubit lifetime under a qubit-flip error syndrome instead of photon losses. The lifetime improvement for the states $\ket{000}_P$ and $\ket{111}_P$ has a much better scaling for increasing error times $T_E$, seen in Fig.~\ref{fig:threeQubit-summary}(c). This is still not a linear scaling, indicative that there are other residual error rates that become dominant for larger $T_E$, but the improvement is very significant over using individually optimized fixed coupling strengths.

\section{\label{sec:vslq} VSLQ}

\begin{figure}
	\includegraphics[width = 0.485\textwidth]{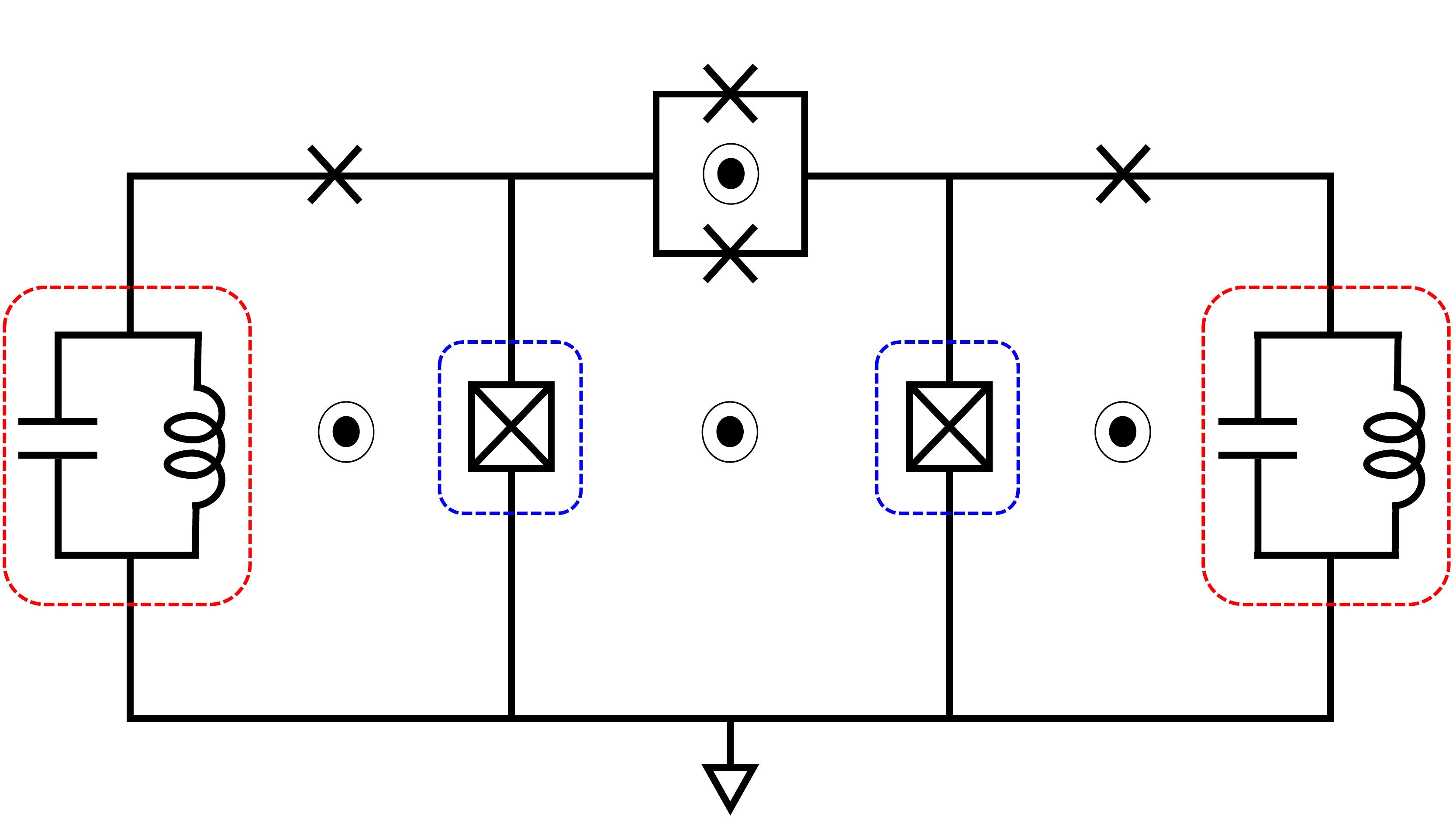}
	\caption{\label{fig:vslq} A possible implementation of the logical qubit, similar to \cite{Kapit2016}. The high-coherence primary qubits are in blue boxes, and the lossy resonators (or qubits) are in red boxes. The approximated rotating wave Hamiltonian [Eq.~(\ref{eq:vslq})] is achieved by modulating the flux drives in the diagram. We refer to the elements of this circuit as shadow left and shadow right ($Sl$ and $Sr$ respectively) for the lossy resonators (red boxes), and simply left and right ($l$ and $r$) for the primary qubits (blue boxes).}
\end{figure}

A simple yet effective architecture, the VSLQ shows great promise in protecting a logical state against single-photon loss errors while suppressing phase errors, all while depending on fully available technology. Having shown the applicability of this pulse-reset technique on a more complicated system, we now apply it to a more realistic implementation, the VSLQ \cite{Kapit2016, Kapit2017}.

\subsection{\label{sec:vslq-system} System}

This circuit consists of two coupled high-coherence qubits each coupled to a lossy qubit or resonator, as shown in Fig.~\ref{fig:vslq}. Following the derivation of the approximated rotating wave Hamiltonian from \cite{Kapit2016}, and using the same notation, we have $H = H_P + H_S + H_{PS}$, with

\begin{equation}\label{eq:vslq}
	\begin{split}
		H_P &= -W\tilde{X}_l\tilde{X}_r + \frac{\delta}{2}\left(P_l^1 + P_r^1\right)\\
		H_S &= \left(W + \frac{\delta}{2}\right)\left(a_{Sl}^{\dagger}a_{Sl} + a_{Sr}^{\dagger}a_{Sr}\right)\\
		H_{PS} &= \Omega\left(a_{l}^{\dagger}a_{Sl}^{\dagger} + a_{r}^{\dagger}a_{Sr}^{\dagger} + \textrm{H.c.}\right),
	\end{split}
\end{equation}
where $\delta$ is a nonlinearity, $\tilde{X}_k = \left(a_k^{\dagger}a_k^{\dagger} + a_ka_k\right)/\sqrt{2}$, $P_k^n = \ketbra{n_k}{n_k},\ k=\{l,r\}$, and we require that for the error correction mechanism to work,
\begin{equation} \label{eq:vslqCondi}
 	\delta \gg W \gg \Omega.
\end{equation}
Here $H_P$ denotes the Hamiltonian for the primary qubits (blue boxes in Fig.~\ref{fig:vslq}), where we only consider the first three levels as the operating space and ignore higher energy states $\left(\ket{3},\ket{4},...\right)$ so as to reduce the size of the Hilbert space and ease numerical simulations. $H_S$ is the Hamiltonian for the ``shadow" resonators (red boxes in Fig.~\ref{fig:vslq}), with $H_{PS}$ the interaction Hamiltonian.

Using the eigenstates for Eq.~(\ref{eq:vslq}), we can define a logical manifold:

\begin{equation}\label{eq:vslq-estates}
	\begin{split}
		\ket{0_L} &= \frac{1}{\sqrt{2}}\left(\ket{0_l} + \ket{2_l}\right) \otimes \frac{1}{\sqrt{2}}\left(\ket{0_r} + \ket{2_r}\right) \otimes \ket{0_{Sl}0_{Sr}}\\
		\ket{1_L} &= \frac{1}{\sqrt{2}}\left(\ket{0_l} - \ket{2_l}\right) \otimes \frac{1}{\sqrt{2}}\left(\ket{0_r} - \ket{2_r}\right) \otimes \ket{0_{Sl}0_{Sr}}.
	\end{split}
\end{equation}
These states are autonomously protected against single-photon losses by the blue-sideband coupling in $H_{PS}$, which is only energetically preferred in the event of a photon loss in one of the primary qubits, leading to long lifetimes of the logical states. This is very similar to the blue-sideband coupling for the single-qubit case shown in Fig.~\ref{fig:singleQubit}(a). A major difference here is that instead of trying to stabilize a single excited state, the goal is to stabilize these superposition states. So while a single photon loss can take the system out of the logical manifold into an error state

\begin{equation}\label{eq:vslq-err}
	\ket{\textrm{Err}_l} = \ket{1_l}\otimes \frac{1}{\sqrt{2}}\left(\ket{0_r} \pm \ket{2_r}\right) \otimes \ket{0_{Sl}0_{Sr}},
\end{equation}
transitions into this state can be induced by $H_{PS}$ for the primary qubits, with the shadows going to $\ket{1_{Sl}0_{Sr}}$. In addition to attempting to minimize these unwanted transitions by $\Omega(t)$, we could also consider transitions into higher energy states. Although highly unlikely, the $\tilde{X}$ operators could induce transitions into $\ket{4}$, which could then stabilize a superposition of $\ket{2} + \ket{4}$, a leakage state. Moreover, accounting for higher excited states would mean consideration of the possibility of the $\tilde{X}$ terms taking an error state $\ket{1}$ into a superposition of $\ket{1} + \ket{3}$. However, given the condition Eq.~(\ref{eq:vslqCondi}), such off-resonant transitions are extremely unlikely due to increasing nonlinearities with increasing energy. Thus for these simulations we limit the primary qubits to three-level systems, and try to address the much more likely transition $\ket{0}+\ket{2} \rightarrow \ket{1}$. The Lindblad operators are $\left\{a_{Sl},a_{l},a_{r},a_{Sr}\right\}$, and we compare the logical state lifetimes $T_L$ against the single qubit lifetime $T_1 = 1/\Gamma_P$, with $\Gamma_P$ being the photon loss rate for the primary qubits ($\Gamma_S$ for the shadow). We assume that it is significantly more likely this error state will be corrected as opposed to finding a higher energy state.

The optimized pulse shape will correspond to the interaction Hamiltonian

\begin{equation}\label{eq:vslqHamCou}
	\begin{split}
		H_{PS} &= \Omega_x(t) \left(a_{l}^{\dagger}a_{Sl}^{\dagger} + a_{r}^{\dagger}a_{Sr}^{\dagger} + \textrm{H.c.}\right)\\
		& \qquad + \Omega_y(t) i\left(a_{l}^{\dagger}a_{Sl}^{\dagger} + a_{r}^{\dagger}a_{Sr}^{\dagger} - \textrm{H.c.}\right),
	\end{split}
\end{equation}
just like we did for the single- and three-qubit cases, with the same goal of having this pulse shape reduce changes to our target logical states defined in Eq.~(\ref{eq:vslq-estates}) and achieving the target operation

\begin{equation}\label{eq:vslqCorrection}
	\begin{split}
		\ket{1_l} &\otimes \frac{1}{\sqrt{2}}\left(\ket{0_r} \pm \ket{2_r}\right) \otimes \ket{0_{Sl}0_{Sr}}\\
		&\rightarrow \frac{1}{\sqrt{2}}\left(\ket{0_l} + \ket{2_l}\right) \otimes \frac{1}{\sqrt{2}}\left(\ket{0_r} + \ket{2_r}\right) \otimes \ket{1_{Sl}0_{Sr}}
	\end{split}
\end{equation}
with high fidelity. During the reset cycle, where $\Gamma_S \gg \Gamma_P$, the lossy qubit returns to its ground state $\ket{1_{Sl}0_{Sr}}\rightarrow \ket{0_{Sl}0_{Sr}}$, returning the state to the logical manifold defined Eq.~(\ref{eq:vslq-estates}).
\begin{figure}
	\includegraphics[width=0.485\textwidth]{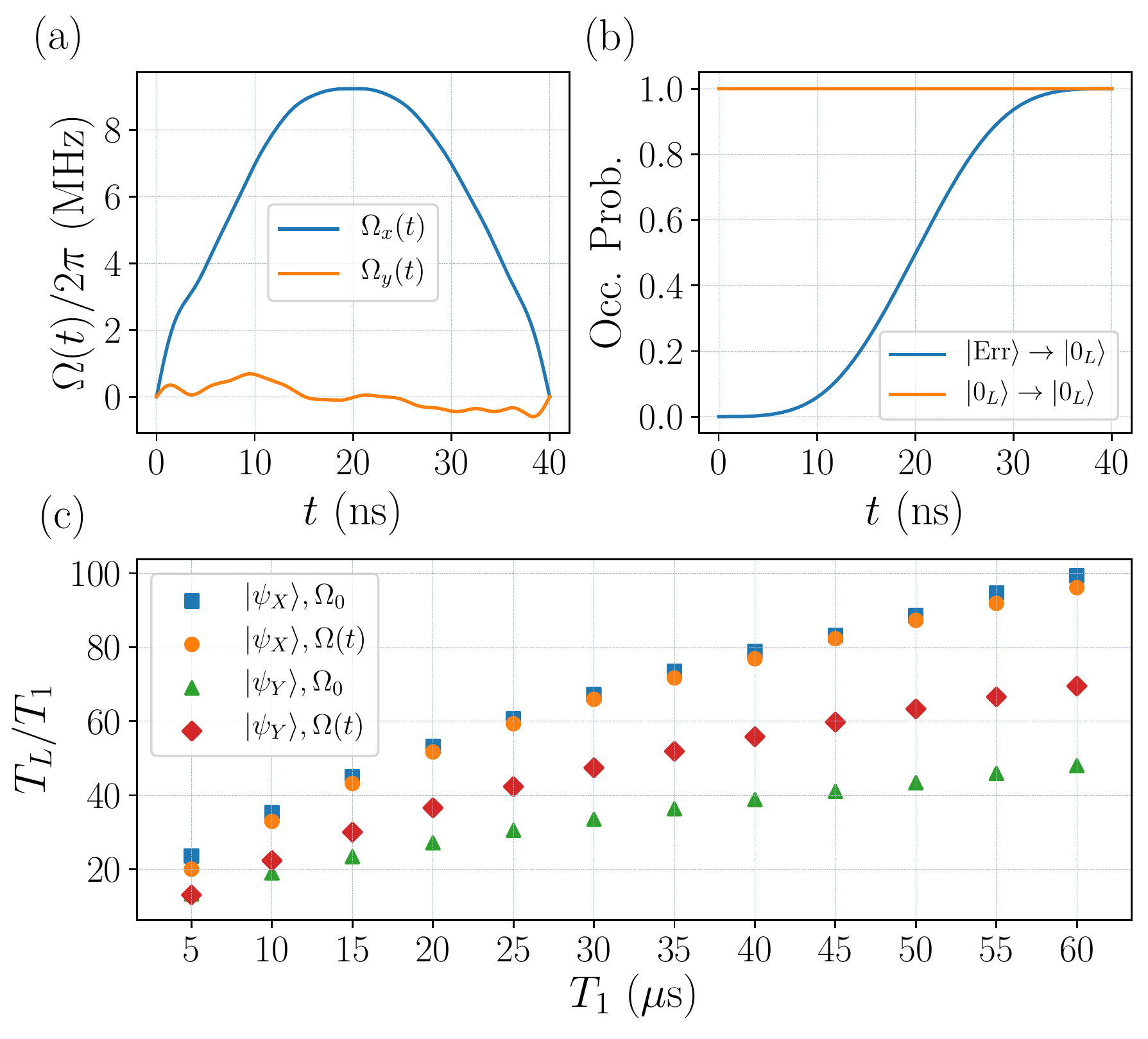}
	\caption{\label{fig:vslq-summary} Results for the VSLQ. (a) Optimized $\Omega_x(t)$ and $\Omega_y(t)$ pulse shapes. (b) High-fidelity target operation of 0.99991 while leaving the target states unchanged. (c) We see the effects of lifetimes for $X_L$ eigenstates and $Y_L$ eigenstates, using definitions from \cite{Kapit2016}. Blue and green are the improvement factors using fixed operating parameters (Table~\ref{tab:vslq-fixed}) for the $X_L$ and $Y_L$ eigenstates, respectively, while orange and red are the improvement factors using pulse-reset cycles. Again, (a) and (b) do not include decoherence, while (c) does.}
\end{figure}

\subsection{\label{sec:vslq-results} Results}

These results using pulse-reset cycles are summarized in Fig.~\ref{fig:vslq-summary}. We compare these to results from running a gradient ascent over the fixed parameter space $\{ \Omega, \omega_S, \Gamma_S \}$, summarized in Table~\ref{tab:vslq-fixed}. These fixed parameter results are our benchmark in looking for an improvement from our pulse-reset protocol.

\begin{figure}
	\includegraphics[width=0.485\textwidth]{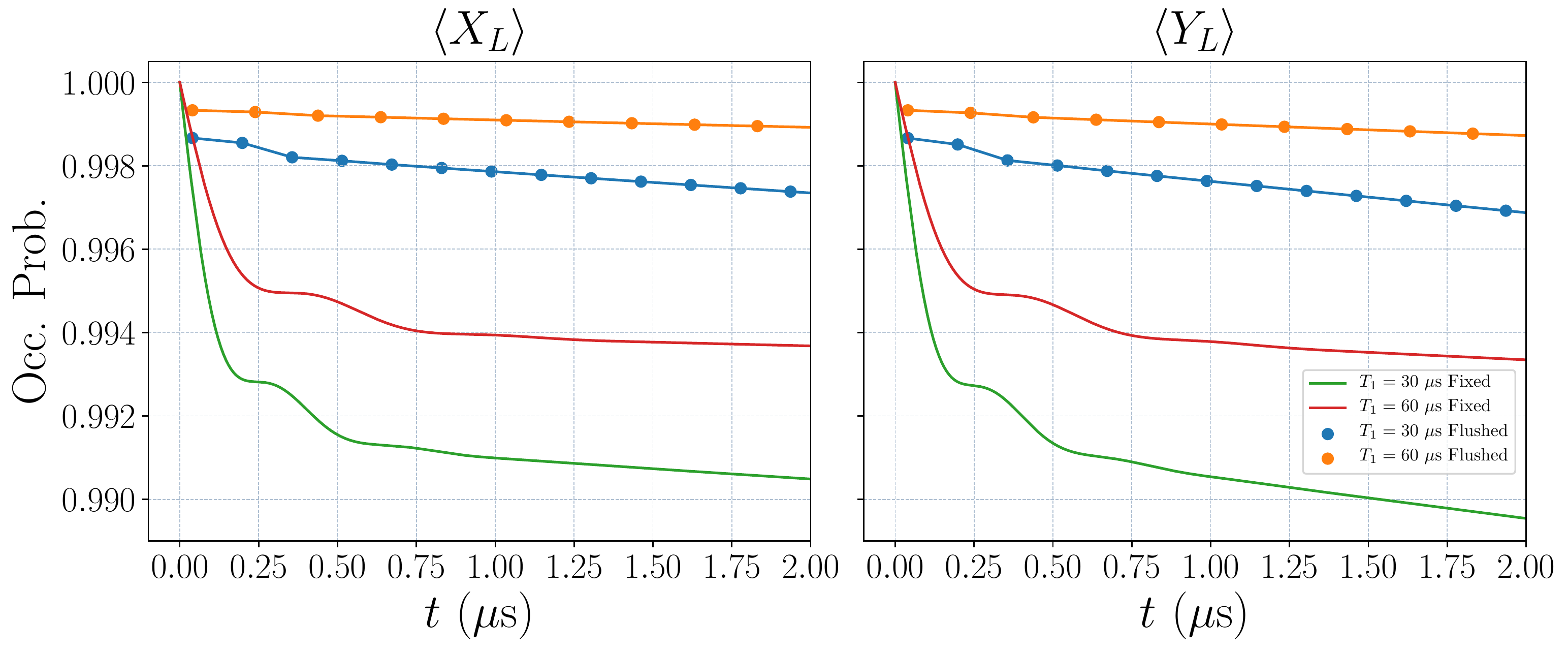}
	\caption{\label{fig:vslq-EVOS} Short-time evolution for the VSLQ $X_L$ eigenstates (left) and $Y_L$ eigenstates (right). We compare using fixed parameters from Table~\ref{tab:vslq-fixed} versus using pulse-reset (flushed) cycles for $T_1=30, 60\ \mu$s.}
\end{figure}

\begin{table}
	\caption{\label{tab:vslq-fixed}
	Gradient ascent results over the fixed parameter space of the VSLQ with $W=2\pi\times35$ MHz, and $\delta = 2\pi\times 350$ MHz. Note the lossy qubit energy approaches the energy from Eq.~(\ref{eq:vslq}), $\omega_S = W+\delta /2 = 2\pi\times210$ MHz.}
	\begin{ruledtabular}
		\begin{tabular}{cccccc}
			$T_1$& $\Omega/2\pi$& $\Gamma_S$& $\omega_S/2\pi$& $T_{X}$& $T_{Y}$\\
			\colrule
			$\mu$s & MHz & $\mu\textrm{s}^{-1}$ & MHz & $\mu$s & $\mu$s \\
			\colrule

			5 & 2.94 & 24.66 & 209.75 & 117 & 66\\
			10 & 2.15 & 18.40 & 209.83 & 353 & 189\\
			15 & 1.81 & 15.32 & 209.90 & 675 & 350\\
			20 & 1.59 & 13.26 & 209.92 & 1061 & 542\\
			25 & 1.43 & 12.09 & 209.94 & 1514 & 762\\
			30 & 1.31 & 11.09 & 209.95 & 2016 & 1005\\
			35 & 1.22 & 10.30 & 209.96 & 2571 & 1271\\
			40 & 1.14 & 9.67 & 209.96 & 3151 & 1553\\
			45 & 1.08 & 9.15 & 209.96 & 3743 & 1846\\
			50 & 1.02 & 8.73 & 209.97 & 4422 & 2168\\
			55 & 0.98 & 8.38 & 209.97 & 5207 & 2524\\
			60 & 0.93 & 8.04 & 209.97 & 5955 & 2879\\
		\end{tabular}
	\end{ruledtabular}
\end{table}

Using the optimized pulse shapes for $\Omega_x(t)$ and $\Omega_y(t)$ from Fig.~\ref{fig:vslq-summary}(a), we achieve an operation fidelity of $1-F < 10^{-4}$, without any noticeable transitions out of the target state $\ket{0_L}$ [Fig.~\ref{fig:vslq-summary}(b)]. We can see that this technique does not show an improvement for the long-time $X_L$ eigenstate lifetimes over using individually, optimized fixed parameters [Fig.~\ref{fig:vslq-summary}(c)]. However, the improvement in the lifetimes for the $Y_L$ eigenstates is very evident. Moreover, the pulse-reset evolution shows a significant improvement over fixed parameters for short-time evolutions, as shown in Fig.~\ref{fig:vslq-EVOS}. Here, we see that for both $X_L$ and $Y_L$ eigenstates, there is a very noticeable advantage, which will prove significantly important for gate operations on the device. In using $\Omega(t)$, we also scan over different reset times for each $T_1$, since each will have a different probability of having a photon in the lossy qubit after the pulse duration $t_p=40$ ns. The loss rate for this lossy qubit is $\Gamma_S = 35\ \mu\textrm{s}^{-1}$ for all different $T_1$ \textit{during} the reset cycle, and again, $\Gamma_S = \Gamma_P$ during the pulse cycles.

While there is a general lack of improvement for the $X_L$ eigenstate lifetimes, this is not the case for the $Y_L$ eigenstates, where we use the definitions from \cite{Kapit2016} to define $Y_L = iX_LZ_L$, with $X_L=\tilde{X}_l$ or $\tilde{X}_r$, and $Z_L=\tilde{Z}_l\tilde{Z}_r$ with $\tilde{Z}_k=P_k^2-P_k^0$. We see in Fig.~\ref{fig:vslq-summary}(c) that there is a clear advantage in the $Y_L$ lifetimes using pulse-reset cycles over fixed operating parameters. While we still expected to see an advantage for the $X_L$ eigenstates as well, we may be seeing these results because of the $Y_L$ eigenstates' sensitivity to more error channels than the $X_L$ eigenstates.


To see this more clearly, consider the $X_L$ eigenstate $\ket{0_L}$ from Eq.~(\ref{eq:vslq-estates}) undergoing a photon loss in the left qubit, taking the system to the error state in Eq.~(\ref{eq:vslq-err}). Photon loss errors in the device are left-right symmetric. For this example we consider an error in the left qubit first only for simplicity, but the same conclusion would be reached starting with a photon loss in the right qubit. As previously explained, this state can then either be corrected back to the logical manifold, or undergo a second photon loss with lower probability on either qubit. In the event of a second photon loss occurring in the right qubit, then the system will be taken to the $\ket{1_l1_r0_{Sl}0_{Sr}}$ state, which is uncorrectable with autonomous error correction having lost all information from the original state. This is considered a logical leakage state for the VSLQ. However, if the second photon loss occurs in the left qubit, then this can take the system to either state:
\begin{equation}
	\begin{split}
	\ket{\psi_1} &= \frac{1}{\sqrt{2}}\left(\ket{0_l} - \ket{2_l}\right) \otimes \frac{1}{\sqrt{2}}\left(\ket{0_r} + \ket{2_r}\right) \otimes \ket{0_{Sl}0_{Sr}},\ \textrm{or}\\ 
	\ket{\psi_2} &= \frac{1}{\sqrt{2}}\left(\ket{0_l} + \ket{2_l}\right) \otimes \frac{1}{\sqrt{2}}\left(\ket{0_r} + \ket{2_r}\right) \otimes \ket{0_{Sl}0_{Sr}}.
	\end{split}
\end{equation}
Having started in $\ket{0_L}$, the first possible outcome from a second loss in the left qubit is an error for any state, while the second is just $\ket{0_L}$, and is therefore not seen as an $X_L$ eigenstate error. However, this last photon loss channel is an error channel for a $Y_L$ eigenstate. And so, this may be why using an optimized target operation for this autonomous mechanism shows a significant improvement in the $Y_L$ lifetimes, but not for the $X_L$ lifetimes.

Regardless of the difference in the long-time performance for the $X_L$ and $Y_L$ eigenstates, we note that the short-time performance of the pulse-reset evolution still outperforms a fixed parameter evolution for either eigenbasis, as seen in Fig.~\ref{fig:vslq-EVOS}. This suggests a very practical use case for short-time applications. Implementing dissipative engineering in digital error correction codes is a very attractive prospect. For example, in the case of the surface code, each error-correction cycle consists of repeated measurement and resetting in short-time periods. Using the advantage in the short-time behavior of the pulse-reset cycle, and a proposed scheme for gates in small logical qubits (\cite{Kapit2018}), would provide a very relevant improvement in gate fidelity and error detection by incorporating small logical qubits within digital codes. This suggests a much more dramatic improvement in long-term error correction than is suggested in Fig.~\ref{fig:vslq-summary}. Exploring this further will be the focus of future work.

\section{\label{sec:conclusion} Conclusion}

We have demonstrated the use of time-parameterized coupling strength and variable loss rates in engineered dissipation for higher-fidelity state stabilization. We accomplished this by using alternating pulse cycles and reset times as shown in Fig.~\ref{fig:pulseReset}. This showed very noticeable advantages for an idealized single-qubit stabilization scheme and for a more complex bit-flip error-correction code. However, for the VSLQ we saw a mixture of results, where the pulse-reset evolutions showed a very clear advantage for the VSLQ eigenstates over short-time evolutions, while giving comparable results as fixed coupling for the long-time evolutions of the VSLQ's $X_L$ eigenstates. Moreover, we did see a clear advantage in prolonging the logical lifetimes of $Y_L$ eigenstates, all of which is a great showcase for the use of this technique for a promising architecture, especially with gate implementations.

Nevertheless, while using autonomous error correction achieves significantly longer lifetimes over their component qubits for superconducting architectures, it comes at the cost of increased complexity. This is something that will need to be carefully characterized for further implementations of small logical qubits in larger quantum computing systems, since physically implementing 2-qubit gates for small logical qubits will in itself be a complex task. However, we have shown that using numerical pulse-shaping can help us achieve highly optimized target operations. For the VSLQ, this could be used to convert leakage errors into logical errors, which are then correctable with dissipative engineering. A leakage state $\ket{1_l1_r0_{Sl}0_{Sr}}$ is both induced, and uncorrectable, by autonomous error correction using fixed operating parameters. However, using a pulse-reset technique with time-varying coupling optimized to perform a target operation taking either qubit from $\ket{1}$ to $\ket{0} + \ket{2}$ has the potential to be used for correcting $\ket{1_l1_r0_{Sl}0_{Sr}}$. This ability to induce transitions from a leakage error state to an ordinary logical error state for the VSLQ would be a key function if ever incorporating small logical qubits into digital error-correction codes.

Further research will be needed to fully understand other prevalent error channels for small logical qubit architectures---especially those involving higher energy states in primary high-coherence qubit devices---as well as limitations of this pulse-reset technique. We would also like to explore whether this technique would provide any sort of advantage in gates for small logical qubits \cite{Kapit2018}, an important characterization for further implementation of these qubits towards fault-tolerance.

\section{\label{sec:acknowledgements} Acknowledgements}

We'd like to thank Nick Materise and Zhijie Tang for the many conversations that helped advance this project. This work was supported by the NSF grant (PHY-1653820) and ARO grant No. W911NF-18-1-0125. It was made possible by the high performance computing resources from the Tulane University Cypress platform, and the Colorado School of Mines Wendian platform.

\bibliography{finalv2}

\begin{thebibliography}{42}%
\makeatletter
\providecommand \@ifxundefined [1]{%
 \@ifx{#1\undefined}
}%
\providecommand \@ifnum [1]{%
 \ifnum #1\expandafter \@firstoftwo
 \else \expandafter \@secondoftwo
 \fi
}%
\providecommand \@ifx [1]{%
 \ifx #1\expandafter \@firstoftwo
 \else \expandafter \@secondoftwo
 \fi
}%
\providecommand \natexlab [1]{#1}%
\providecommand \enquote  [1]{``#1''}%
\providecommand \bibnamefont  [1]{#1}%
\providecommand \bibfnamefont [1]{#1}%
\providecommand \citenamefont [1]{#1}%
\providecommand \href@noop [0]{\@secondoftwo}%
\providecommand \href [0]{\begingroup \@sanitize@url \@href}%
\providecommand \@href[1]{\@@startlink{#1}\@@href}%
\providecommand \@@href[1]{\endgroup#1\@@endlink}%
\providecommand \@sanitize@url [0]{\catcode `\\12\catcode `\$12\catcode
  `\&12\catcode `\#12\catcode `\^12\catcode `\_12\catcode `\%12\relax}%
\providecommand \@@startlink[1]{}%
\providecommand \@@endlink[0]{}%
\providecommand \url  [0]{\begingroup\@sanitize@url \@url }%
\providecommand \@url [1]{\endgroup\@href {#1}{\urlprefix }}%
\providecommand \urlprefix  [0]{URL }%
\providecommand \Eprint [0]{\href }%
\providecommand \doibase [0]{https://doi.org/}%
\providecommand \selectlanguage [0]{\@gobble}%
\providecommand \bibinfo  [0]{\@secondoftwo}%
\providecommand \bibfield  [0]{\@secondoftwo}%
\providecommand \translation [1]{[#1]}%
\providecommand \BibitemOpen [0]{}%
\providecommand \bibitemStop [0]{}%
\providecommand \bibitemNoStop [0]{.\EOS\space}%
\providecommand \EOS [0]{\spacefactor3000\relax}%
\providecommand \BibitemShut  [1]{\csname bibitem#1\endcsname}%
\let\auto@bib@innerbib\@empty
\bibitem [{\citenamefont {Fowler}\ \emph {et~al.}(2012)\citenamefont {Fowler},
  \citenamefont {Mariantoni}, \citenamefont {Martinis},\ and\ \citenamefont
  {Cleland}}]{Fowler2012}%
  \BibitemOpen
  \bibfield  {author} {\bibinfo {author} {\bibfnamefont {A.~G.}\ \bibnamefont
  {Fowler}}, \bibinfo {author} {\bibfnamefont {M.}~\bibnamefont {Mariantoni}},
  \bibinfo {author} {\bibfnamefont {J.~M.}\ \bibnamefont {Martinis}},\ and\
  \bibinfo {author} {\bibfnamefont {A.~N.}\ \bibnamefont {Cleland}},\
  }\bibfield  {title} {\bibinfo {title} {{Surface codes: Towards practical
  large-scale quantum computation}},\ }\href@noop {} {\bibfield  {journal}
  {\bibinfo  {journal} {Physical Review A - Atomic, Molecular, and Optical
  Physics}\ }\textbf {\bibinfo {volume} {86}} (\bibinfo {year}
  {2012})}\BibitemShut {NoStop}%
\bibitem [{\citenamefont {Kapit}(2017)}]{Kapit2017}%
  \BibitemOpen
  \bibfield  {author} {\bibinfo {author} {\bibfnamefont {E.}~\bibnamefont
  {Kapit}},\ }\bibfield  {title} {\bibinfo {title} {{The upside of noise:
  engineered dissipation as a resource in superconducting circuits}},\ }\href
  {https://doi.org/10.1088/2058-9565/aa7e5d} {\bibfield  {journal} {\bibinfo
  {journal} {Quantum Science and Technology}\ }\textbf {\bibinfo {volume}
  {2}},\ \bibinfo {pages} {033002} (\bibinfo {year} {2017})}\BibitemShut
  {NoStop}%
\bibitem [{\citenamefont {Ma}\ \emph {et~al.}(2020)\citenamefont {Ma},
  \citenamefont {Xu}, \citenamefont {Mu}, \citenamefont {Cai}, \citenamefont
  {Hu}, \citenamefont {Wang}, \citenamefont {Pan}, \citenamefont {Wang},
  \citenamefont {Song}, \citenamefont {Zou},\ and\ \citenamefont
  {Sun}}]{Ma2019}%
  \BibitemOpen
  \bibfield  {author} {\bibinfo {author} {\bibfnamefont {Y.}~\bibnamefont
  {Ma}}, \bibinfo {author} {\bibfnamefont {Y.}~\bibnamefont {Xu}}, \bibinfo
  {author} {\bibfnamefont {X.}~\bibnamefont {Mu}}, \bibinfo {author}
  {\bibfnamefont {W.}~\bibnamefont {Cai}}, \bibinfo {author} {\bibfnamefont
  {L.}~\bibnamefont {Hu}}, \bibinfo {author} {\bibfnamefont {W.}~\bibnamefont
  {Wang}}, \bibinfo {author} {\bibfnamefont {X.}~\bibnamefont {Pan}}, \bibinfo
  {author} {\bibfnamefont {H.}~\bibnamefont {Wang}}, \bibinfo {author}
  {\bibfnamefont {Y.~P.}\ \bibnamefont {Song}}, \bibinfo {author}
  {\bibfnamefont {C.~L.}\ \bibnamefont {Zou}},\ and\ \bibinfo {author}
  {\bibfnamefont {L.}~\bibnamefont {Sun}},\ }\bibfield  {title} {\bibinfo
  {title} {{Error-transparent operations on a logical qubit protected by
  quantum error correction}},\ }\href {http://arxiv.org/abs/1909.06803}
  {\bibfield  {journal} {\bibinfo  {journal} {Nature Physics}\ } (\bibinfo
  {year} {2020})}\BibitemShut {NoStop}%
\bibitem [{\citenamefont {Gertler}\ \emph {et~al.}(2020)\citenamefont
  {Gertler}, \citenamefont {Baker}, \citenamefont {Li}, \citenamefont {Shirol},
  \citenamefont {Koch},\ and\ \citenamefont {Wang}}]{Gertler2020}%
  \BibitemOpen
  \bibfield  {author} {\bibinfo {author} {\bibfnamefont {J.~M.}\ \bibnamefont
  {Gertler}}, \bibinfo {author} {\bibfnamefont {B.}~\bibnamefont {Baker}},
  \bibinfo {author} {\bibfnamefont {J.}~\bibnamefont {Li}}, \bibinfo {author}
  {\bibfnamefont {S.}~\bibnamefont {Shirol}}, \bibinfo {author} {\bibfnamefont
  {J.}~\bibnamefont {Koch}},\ and\ \bibinfo {author} {\bibfnamefont
  {C.}~\bibnamefont {Wang}},\ }\bibfield  {title} {\bibinfo {title}
  {{Protecting a Bosonic Qubit with Autonomous Quantum Error Correction}},\
  }\href {http://arxiv.org/abs/2004.09322} {\  (\bibinfo {year} {2020})},\
  \Eprint {https://arxiv.org/abs/2004.09322} {arXiv:2004.09322} \BibitemShut
  {NoStop}%
\bibitem [{\citenamefont {Kristensen}\ \emph {et~al.}(2019)\citenamefont
  {Kristensen}, \citenamefont {Kjaergaard}, \citenamefont {Andersen},\ and\
  \citenamefont {Zinner}}]{Kristensen2019}%
  \BibitemOpen
  \bibfield  {author} {\bibinfo {author} {\bibfnamefont {L.~B.}\ \bibnamefont
  {Kristensen}}, \bibinfo {author} {\bibfnamefont {M.}~\bibnamefont
  {Kjaergaard}}, \bibinfo {author} {\bibfnamefont {C.~K.}\ \bibnamefont
  {Andersen}},\ and\ \bibinfo {author} {\bibfnamefont {N.~T.}\ \bibnamefont
  {Zinner}},\ }\bibfield  {title} {\bibinfo {title} {{Hybrid Quantum Error
  Correction in Qubit Architectures}},\ }\href
  {http://arxiv.org/abs/1909.09112} {\  (\bibinfo {year} {2019})},\ \Eprint
  {https://arxiv.org/abs/1909.09112} {arXiv:1909.09112} \BibitemShut {NoStop}%
\bibitem [{\citenamefont {Leghtas}\ \emph {et~al.}(2013)\citenamefont
  {Leghtas}, \citenamefont {Kirchmair}, \citenamefont {Vlastakis},
  \citenamefont {Schoelkopf}, \citenamefont {Devoret},\ and\ \citenamefont
  {Mirrahimi}}]{Leghtas2013}%
  \BibitemOpen
  \bibfield  {author} {\bibinfo {author} {\bibfnamefont {Z.}~\bibnamefont
  {Leghtas}}, \bibinfo {author} {\bibfnamefont {G.}~\bibnamefont {Kirchmair}},
  \bibinfo {author} {\bibfnamefont {B.}~\bibnamefont {Vlastakis}}, \bibinfo
  {author} {\bibfnamefont {R.~J.}\ \bibnamefont {Schoelkopf}}, \bibinfo
  {author} {\bibfnamefont {M.~H.}\ \bibnamefont {Devoret}},\ and\ \bibinfo
  {author} {\bibfnamefont {M.}~\bibnamefont {Mirrahimi}},\ }\bibfield  {title}
  {\bibinfo {title} {{Hardware-Efficient Autonomous Quantum Memory
  Protection}},\ }\href {https://doi.org/10.1103/PhysRevLett.111.120501}
  {\bibfield  {journal} {\bibinfo  {journal} {Physical Review Letters}\
  }\textbf {\bibinfo {volume} {111}},\ \bibinfo {pages} {120501} (\bibinfo
  {year} {2013})}\BibitemShut {NoStop}%
\bibitem [{\citenamefont {Mirrahimi}\ \emph {et~al.}(2014)\citenamefont
  {Mirrahimi}, \citenamefont {Leghtas}, \citenamefont {Albert}, \citenamefont
  {Touzard}, \citenamefont {Schoelkopf}, \citenamefont {Jiang},\ and\
  \citenamefont {Devoret}}]{Mirrahimi2014}%
  \BibitemOpen
  \bibfield  {author} {\bibinfo {author} {\bibfnamefont {M.}~\bibnamefont
  {Mirrahimi}}, \bibinfo {author} {\bibfnamefont {Z.}~\bibnamefont {Leghtas}},
  \bibinfo {author} {\bibfnamefont {V.~V.}\ \bibnamefont {Albert}}, \bibinfo
  {author} {\bibfnamefont {S.}~\bibnamefont {Touzard}}, \bibinfo {author}
  {\bibfnamefont {R.~J.}\ \bibnamefont {Schoelkopf}}, \bibinfo {author}
  {\bibfnamefont {L.}~\bibnamefont {Jiang}},\ and\ \bibinfo {author}
  {\bibfnamefont {M.~H.}\ \bibnamefont {Devoret}},\ }\bibfield  {title}
  {\bibinfo {title} {{Dynamically protected cat-qubits: a new paradigm for
  universal quantum computation}},\ }\href
  {https://doi.org/10.1088/1367-2630/16/4/045014} {\bibfield  {journal}
  {\bibinfo  {journal} {New Journal of Physics}\ }\textbf {\bibinfo {volume}
  {16}},\ \bibinfo {pages} {045014} (\bibinfo {year} {2014})}\BibitemShut
  {NoStop}%
\bibitem [{\citenamefont {Sun}\ \emph {et~al.}(2014)\citenamefont {Sun},
  \citenamefont {Petrenko}, \citenamefont {Leghtas}, \citenamefont {Vlastakis},
  \citenamefont {Kirchmair}, \citenamefont {Sliwa}, \citenamefont {Narla},
  \citenamefont {Hatridge}, \citenamefont {Shankar}, \citenamefont {Blumoff},
  \citenamefont {Frunzio}, \citenamefont {Mirrahimi}, \citenamefont {Devoret},\
  and\ \citenamefont {Schoelkopf}}]{Sun2014}%
  \BibitemOpen
  \bibfield  {author} {\bibinfo {author} {\bibfnamefont {L.}~\bibnamefont
  {Sun}}, \bibinfo {author} {\bibfnamefont {A.}~\bibnamefont {Petrenko}},
  \bibinfo {author} {\bibfnamefont {Z.}~\bibnamefont {Leghtas}}, \bibinfo
  {author} {\bibfnamefont {B.}~\bibnamefont {Vlastakis}}, \bibinfo {author}
  {\bibfnamefont {G.}~\bibnamefont {Kirchmair}}, \bibinfo {author}
  {\bibfnamefont {K.~M.}\ \bibnamefont {Sliwa}}, \bibinfo {author}
  {\bibfnamefont {A.}~\bibnamefont {Narla}}, \bibinfo {author} {\bibfnamefont
  {M.}~\bibnamefont {Hatridge}}, \bibinfo {author} {\bibfnamefont
  {S.}~\bibnamefont {Shankar}}, \bibinfo {author} {\bibfnamefont
  {J.}~\bibnamefont {Blumoff}}, \bibinfo {author} {\bibfnamefont
  {L.}~\bibnamefont {Frunzio}}, \bibinfo {author} {\bibfnamefont
  {M.}~\bibnamefont {Mirrahimi}}, \bibinfo {author} {\bibfnamefont {M.~H.}\
  \bibnamefont {Devoret}},\ and\ \bibinfo {author} {\bibfnamefont {R.~J.}\
  \bibnamefont {Schoelkopf}},\ }\bibfield  {title} {\bibinfo {title} {{Tracking
  photon jumps with repeated quantum non-demolition parity measurements}},\
  }\href {https://doi.org/10.1038/nature13436} {\bibfield  {journal} {\bibinfo
  {journal} {Nature}\ }\textbf {\bibinfo {volume} {511}},\ \bibinfo {pages}
  {444} (\bibinfo {year} {2014})}\BibitemShut {NoStop}%
\bibitem [{\citenamefont {Leghtas}\ \emph {et~al.}(2015)\citenamefont
  {Leghtas}, \citenamefont {Touzard}, \citenamefont {Pop}, \citenamefont {Kou},
  \citenamefont {Vlastakis}, \citenamefont {Petrenko}, \citenamefont {Sliwa},
  \citenamefont {Narla}, \citenamefont {Shankar}, \citenamefont {Hatridge},
  \citenamefont {Reagor}, \citenamefont {Frunzio}, \citenamefont {Schoelkopf},
  \citenamefont {Mirrahimi},\ and\ \citenamefont {Devoret}}]{Leghtas2015}%
  \BibitemOpen
  \bibfield  {author} {\bibinfo {author} {\bibfnamefont {Z.}~\bibnamefont
  {Leghtas}}, \bibinfo {author} {\bibfnamefont {S.}~\bibnamefont {Touzard}},
  \bibinfo {author} {\bibfnamefont {I.~M.}\ \bibnamefont {Pop}}, \bibinfo
  {author} {\bibfnamefont {A.}~\bibnamefont {Kou}}, \bibinfo {author}
  {\bibfnamefont {B.}~\bibnamefont {Vlastakis}}, \bibinfo {author}
  {\bibfnamefont {A.}~\bibnamefont {Petrenko}}, \bibinfo {author}
  {\bibfnamefont {K.~M.}\ \bibnamefont {Sliwa}}, \bibinfo {author}
  {\bibfnamefont {A.}~\bibnamefont {Narla}}, \bibinfo {author} {\bibfnamefont
  {S.}~\bibnamefont {Shankar}}, \bibinfo {author} {\bibfnamefont {M.~J.}\
  \bibnamefont {Hatridge}}, \bibinfo {author} {\bibfnamefont {M.}~\bibnamefont
  {Reagor}}, \bibinfo {author} {\bibfnamefont {L.}~\bibnamefont {Frunzio}},
  \bibinfo {author} {\bibfnamefont {R.~J.}\ \bibnamefont {Schoelkopf}},
  \bibinfo {author} {\bibfnamefont {M.}~\bibnamefont {Mirrahimi}},\ and\
  \bibinfo {author} {\bibfnamefont {M.~H.}\ \bibnamefont {Devoret}},\
  }\bibfield  {title} {\bibinfo {title} {{Quantum engineering. Confining the
  state of light to a quantum manifold by engineered two-photon loss.}},\
  }\href {https://doi.org/10.1126/science.aaa2085} {\bibfield  {journal}
  {\bibinfo  {journal} {Science (New York, N.Y.)}\ }\textbf {\bibinfo {volume}
  {347}},\ \bibinfo {pages} {853} (\bibinfo {year} {2015})}\BibitemShut
  {NoStop}%
\bibitem [{\citenamefont {Albert}\ \emph {et~al.}(2016)\citenamefont {Albert},
  \citenamefont {Shu}, \citenamefont {Krastanov}, \citenamefont {Shen},
  \citenamefont {Liu}, \citenamefont {Yang}, \citenamefont {Schoelkopf},
  \citenamefont {Mirrahimi}, \citenamefont {Devoret},\ and\ \citenamefont
  {Jiang}}]{Albert2016}%
  \BibitemOpen
  \bibfield  {author} {\bibinfo {author} {\bibfnamefont {V.~V.}\ \bibnamefont
  {Albert}}, \bibinfo {author} {\bibfnamefont {C.}~\bibnamefont {Shu}},
  \bibinfo {author} {\bibfnamefont {S.}~\bibnamefont {Krastanov}}, \bibinfo
  {author} {\bibfnamefont {C.}~\bibnamefont {Shen}}, \bibinfo {author}
  {\bibfnamefont {R.-B.}\ \bibnamefont {Liu}}, \bibinfo {author} {\bibfnamefont
  {Z.-B.}\ \bibnamefont {Yang}}, \bibinfo {author} {\bibfnamefont {R.~J.}\
  \bibnamefont {Schoelkopf}}, \bibinfo {author} {\bibfnamefont
  {M.}~\bibnamefont {Mirrahimi}}, \bibinfo {author} {\bibfnamefont {M.~H.}\
  \bibnamefont {Devoret}},\ and\ \bibinfo {author} {\bibfnamefont
  {L.}~\bibnamefont {Jiang}},\ }\bibfield  {title} {\bibinfo {title}
  {{Holonomic Quantum Control with Continuous Variable Systems}},\ }\href
  {https://doi.org/10.1103/PhysRevLett.116.140502} {\bibfield  {journal}
  {\bibinfo  {journal} {Physical Review Letters}\ }\textbf {\bibinfo {volume}
  {116}},\ \bibinfo {pages} {140502} (\bibinfo {year} {2016})}\BibitemShut
  {NoStop}%
\bibitem [{\citenamefont {Ofek}\ \emph {et~al.}(2016)\citenamefont {Ofek},
  \citenamefont {Petrenko}, \citenamefont {Heeres}, \citenamefont {Reinhold},
  \citenamefont {Leghtas}, \citenamefont {Vlastakis}, \citenamefont {Liu},
  \citenamefont {Frunzio}, \citenamefont {Girvin}, \citenamefont {Jiang},
  \citenamefont {Mirrahimi}, \citenamefont {Devoret},\ and\ \citenamefont
  {Schoelkopf}}]{Ofek2016}%
  \BibitemOpen
  \bibfield  {author} {\bibinfo {author} {\bibfnamefont {N.}~\bibnamefont
  {Ofek}}, \bibinfo {author} {\bibfnamefont {A.}~\bibnamefont {Petrenko}},
  \bibinfo {author} {\bibfnamefont {R.}~\bibnamefont {Heeres}}, \bibinfo
  {author} {\bibfnamefont {P.}~\bibnamefont {Reinhold}}, \bibinfo {author}
  {\bibfnamefont {Z.}~\bibnamefont {Leghtas}}, \bibinfo {author} {\bibfnamefont
  {B.}~\bibnamefont {Vlastakis}}, \bibinfo {author} {\bibfnamefont
  {Y.}~\bibnamefont {Liu}}, \bibinfo {author} {\bibfnamefont {L.}~\bibnamefont
  {Frunzio}}, \bibinfo {author} {\bibfnamefont {S.~M.}\ \bibnamefont {Girvin}},
  \bibinfo {author} {\bibfnamefont {L.}~\bibnamefont {Jiang}}, \bibinfo
  {author} {\bibfnamefont {M.}~\bibnamefont {Mirrahimi}}, \bibinfo {author}
  {\bibfnamefont {M.~H.}\ \bibnamefont {Devoret}},\ and\ \bibinfo {author}
  {\bibfnamefont {R.~J.}\ \bibnamefont {Schoelkopf}},\ }\bibfield  {title}
  {\bibinfo {title} {{Extending the lifetime of a quantum bit with error
  correction in superconducting circuits}},\ }\href
  {https://doi.org/10.1038/nature18949} {\bibfield  {journal} {\bibinfo
  {journal} {Nature}\ }\textbf {\bibinfo {volume} {536}},\ \bibinfo {pages}
  {441} (\bibinfo {year} {2016})}\BibitemShut {NoStop}%
\bibitem [{\citenamefont {Cohen}\ \emph {et~al.}(2017)\citenamefont {Cohen},
  \citenamefont {Smith}, \citenamefont {Devoret},\ and\ \citenamefont
  {Mirrahimi}}]{Cohen2017}%
  \BibitemOpen
  \bibfield  {author} {\bibinfo {author} {\bibfnamefont {J.}~\bibnamefont
  {Cohen}}, \bibinfo {author} {\bibfnamefont {W.~C.}\ \bibnamefont {Smith}},
  \bibinfo {author} {\bibfnamefont {M.~H.}\ \bibnamefont {Devoret}},\ and\
  \bibinfo {author} {\bibfnamefont {M.}~\bibnamefont {Mirrahimi}},\ }\bibfield
  {title} {\bibinfo {title} {{Degeneracy-Preserving Quantum Nondemolition
  Measurement of Parity-Type Observables for Cat Qubits}},\ }\href
  {https://doi.org/10.1103/PhysRevLett.119.060503} {\bibfield  {journal}
  {\bibinfo  {journal} {Physical Review Letters}\ }\textbf {\bibinfo {volume}
  {119}},\ \bibinfo {pages} {060503} (\bibinfo {year} {2017})}\BibitemShut
  {NoStop}%
\bibitem [{\citenamefont {Mundhada}\ \emph {et~al.}(2017)\citenamefont
  {Mundhada}, \citenamefont {Grimm}, \citenamefont {Touzard}, \citenamefont
  {Vool}, \citenamefont {Shankar}, \citenamefont {Devoret},\ and\ \citenamefont
  {Mirrahimi}}]{Mundhada2017}%
  \BibitemOpen
  \bibfield  {author} {\bibinfo {author} {\bibfnamefont {S.~O.}\ \bibnamefont
  {Mundhada}}, \bibinfo {author} {\bibfnamefont {A.}~\bibnamefont {Grimm}},
  \bibinfo {author} {\bibfnamefont {S.}~\bibnamefont {Touzard}}, \bibinfo
  {author} {\bibfnamefont {U.}~\bibnamefont {Vool}}, \bibinfo {author}
  {\bibfnamefont {S.}~\bibnamefont {Shankar}}, \bibinfo {author} {\bibfnamefont
  {M.~H.}\ \bibnamefont {Devoret}},\ and\ \bibinfo {author} {\bibfnamefont
  {M.}~\bibnamefont {Mirrahimi}},\ }\bibfield  {title} {\bibinfo {title}
  {{Generating higher-order quantum dissipation from lower-order parametric
  processes}},\ }\href {https://doi.org/10.1088/2058-9565/aa6e9d} {\bibfield
  {journal} {\bibinfo  {journal} {Quantum Science and Technology}\ }\textbf
  {\bibinfo {volume} {2}},\ \bibinfo {pages} {024005} (\bibinfo {year}
  {2017})}\BibitemShut {NoStop}%
\bibitem [{\citenamefont {Michael}\ \emph {et~al.}(2016)\citenamefont
  {Michael}, \citenamefont {Silveri}, \citenamefont {Brierley}, \citenamefont
  {Albert}, \citenamefont {Salmilehto}, \citenamefont {Jiang},\ and\
  \citenamefont {Girvin}}]{Michael2016}%
  \BibitemOpen
  \bibfield  {author} {\bibinfo {author} {\bibfnamefont {M.~H.}\ \bibnamefont
  {Michael}}, \bibinfo {author} {\bibfnamefont {M.}~\bibnamefont {Silveri}},
  \bibinfo {author} {\bibfnamefont {R.}~\bibnamefont {Brierley}}, \bibinfo
  {author} {\bibfnamefont {V.~V.}\ \bibnamefont {Albert}}, \bibinfo {author}
  {\bibfnamefont {J.}~\bibnamefont {Salmilehto}}, \bibinfo {author}
  {\bibfnamefont {L.}~\bibnamefont {Jiang}},\ and\ \bibinfo {author}
  {\bibfnamefont {S.}~\bibnamefont {Girvin}},\ }\bibfield  {title} {\bibinfo
  {title} {{New Class of Quantum Error-Correcting Codes for a Bosonic Mode}},\
  }\href {https://doi.org/10.1103/PhysRevX.6.031006} {\bibfield  {journal}
  {\bibinfo  {journal} {Physical Review X}\ }\textbf {\bibinfo {volume} {6}},\
  \bibinfo {pages} {031006} (\bibinfo {year} {2016})}\BibitemShut {NoStop}%
\bibitem [{\citenamefont {Wang}\ \emph {et~al.}(2016)\citenamefont {Wang},
  \citenamefont {Gao}, \citenamefont {Reinhold}, \citenamefont {Heeres},
  \citenamefont {Ofek}, \citenamefont {Chou}, \citenamefont {Axline},
  \citenamefont {Reagor}, \citenamefont {Blumoff}, \citenamefont {Sliwa},
  \citenamefont {Frunzio}, \citenamefont {Girvin}, \citenamefont {Jiang},
  \citenamefont {Mirrahimi}, \citenamefont {Devoret},\ and\ \citenamefont
  {Schoelkopf}}]{Wang2016}%
  \BibitemOpen
  \bibfield  {author} {\bibinfo {author} {\bibfnamefont {C.}~\bibnamefont
  {Wang}}, \bibinfo {author} {\bibfnamefont {Y.~Y.}\ \bibnamefont {Gao}},
  \bibinfo {author} {\bibfnamefont {P.}~\bibnamefont {Reinhold}}, \bibinfo
  {author} {\bibfnamefont {R.~W.}\ \bibnamefont {Heeres}}, \bibinfo {author}
  {\bibfnamefont {N.}~\bibnamefont {Ofek}}, \bibinfo {author} {\bibfnamefont
  {K.}~\bibnamefont {Chou}}, \bibinfo {author} {\bibfnamefont {C.}~\bibnamefont
  {Axline}}, \bibinfo {author} {\bibfnamefont {M.}~\bibnamefont {Reagor}},
  \bibinfo {author} {\bibfnamefont {J.}~\bibnamefont {Blumoff}}, \bibinfo
  {author} {\bibfnamefont {K.~M.}\ \bibnamefont {Sliwa}}, \bibinfo {author}
  {\bibfnamefont {L.}~\bibnamefont {Frunzio}}, \bibinfo {author} {\bibfnamefont
  {S.~M.}\ \bibnamefont {Girvin}}, \bibinfo {author} {\bibfnamefont
  {L.}~\bibnamefont {Jiang}}, \bibinfo {author} {\bibfnamefont
  {M.}~\bibnamefont {Mirrahimi}}, \bibinfo {author} {\bibfnamefont {M.~H.}\
  \bibnamefont {Devoret}},\ and\ \bibinfo {author} {\bibfnamefont {R.~J.}\
  \bibnamefont {Schoelkopf}},\ }\bibfield  {title} {\bibinfo {title} {{A
  Schr{\"{o}}dinger cat living in two boxes.}},\ }\href
  {https://doi.org/10.1126/science.aaf2941} {\bibfield  {journal} {\bibinfo
  {journal} {Science (New York, N.Y.)}\ }\textbf {\bibinfo {volume} {352}},\
  \bibinfo {pages} {1087} (\bibinfo {year} {2016})}\BibitemShut {NoStop}%
\bibitem [{\citenamefont {Heeres}\ \emph {et~al.}(2017)\citenamefont {Heeres},
  \citenamefont {Reinhold}, \citenamefont {Ofek}, \citenamefont {Frunzio},
  \citenamefont {Jiang}, \citenamefont {Devoret},\ and\ \citenamefont
  {Schoelkopf}}]{Heeres2017}%
  \BibitemOpen
  \bibfield  {author} {\bibinfo {author} {\bibfnamefont {R.~W.}\ \bibnamefont
  {Heeres}}, \bibinfo {author} {\bibfnamefont {P.}~\bibnamefont {Reinhold}},
  \bibinfo {author} {\bibfnamefont {N.}~\bibnamefont {Ofek}}, \bibinfo {author}
  {\bibfnamefont {L.}~\bibnamefont {Frunzio}}, \bibinfo {author} {\bibfnamefont
  {L.}~\bibnamefont {Jiang}}, \bibinfo {author} {\bibfnamefont {M.~H.}\
  \bibnamefont {Devoret}},\ and\ \bibinfo {author} {\bibfnamefont {R.~J.}\
  \bibnamefont {Schoelkopf}},\ }\bibfield  {title} {\bibinfo {title}
  {{Implementing a universal gate set on a logical qubit encoded in an
  oscillator}},\ }\href {https://doi.org/10.1038/s41467-017-00045-1} {\bibfield
   {journal} {\bibinfo  {journal} {Nature Communications}\ }\textbf {\bibinfo
  {volume} {8}},\ \bibinfo {pages} {94} (\bibinfo {year} {2017})}\BibitemShut
  {NoStop}%
\bibitem [{\citenamefont {Puri}\ \emph {et~al.}(2017)\citenamefont {Puri},
  \citenamefont {Boutin},\ and\ \citenamefont {Blais}}]{Puri2017}%
  \BibitemOpen
  \bibfield  {author} {\bibinfo {author} {\bibfnamefont {S.}~\bibnamefont
  {Puri}}, \bibinfo {author} {\bibfnamefont {S.}~\bibnamefont {Boutin}},\ and\
  \bibinfo {author} {\bibfnamefont {A.}~\bibnamefont {Blais}},\ }\bibfield
  {title} {\bibinfo {title} {{Engineering the quantum states of light in a
  Kerr-nonlinear resonator by two-photon driving}},\ }\href
  {https://doi.org/10.1038/s41534-017-0019-1} {\bibfield  {journal} {\bibinfo
  {journal} {npj Quantum Information}\ }\textbf {\bibinfo {volume} {3}},\
  \bibinfo {pages} {18} (\bibinfo {year} {2017})}\BibitemShut {NoStop}%
\bibitem [{\citenamefont {Kapit}(2016)}]{Kapit2016}%
  \BibitemOpen
  \bibfield  {author} {\bibinfo {author} {\bibfnamefont {E.}~\bibnamefont
  {Kapit}},\ }\bibfield  {title} {\bibinfo {title} {{Hardware-Efficient and
  Fully Autonomous Quantum Error Correction in Superconducting Circuits}},\
  }\href {https://doi.org/10.1103/PhysRevLett.116.150501} {\bibfield  {journal}
  {\bibinfo  {journal} {Physical Review Letters}\ }\textbf {\bibinfo {volume}
  {116}},\ \bibinfo {pages} {150501} (\bibinfo {year} {2016})}\BibitemShut
  {NoStop}%
\bibitem [{\citenamefont {Beaudoin}\ \emph {et~al.}(2012)\citenamefont
  {Beaudoin}, \citenamefont {{Da Silva}}, \citenamefont {Dutton},\ and\
  \citenamefont {Blais}}]{Beaudoin2012}%
  \BibitemOpen
  \bibfield  {author} {\bibinfo {author} {\bibfnamefont {F.}~\bibnamefont
  {Beaudoin}}, \bibinfo {author} {\bibfnamefont {M.~P.}\ \bibnamefont {{Da
  Silva}}}, \bibinfo {author} {\bibfnamefont {Z.}~\bibnamefont {Dutton}},\ and\
  \bibinfo {author} {\bibfnamefont {A.}~\bibnamefont {Blais}},\ }\bibfield
  {title} {\bibinfo {title} {{First-order sidebands in circuit QED using qubit
  frequency modulation}},\ }\href {https://doi.org/10.1103/PhysRevA.86.022305}
  {\bibfield  {journal} {\bibinfo  {journal} {Physical Review A - Atomic,
  Molecular, and Optical Physics}\ }\textbf {\bibinfo {volume} {86}},\ \bibinfo
  {pages} {022305} (\bibinfo {year} {2012})}\BibitemShut {NoStop}%
\bibitem [{\citenamefont {Strand}\ \emph {et~al.}(2013)\citenamefont {Strand},
  \citenamefont {Ware}, \citenamefont {Beaudoin}, \citenamefont {Ohki},
  \citenamefont {Johnson}, \citenamefont {Blais},\ and\ \citenamefont
  {Plourde}}]{Strand2013}%
  \BibitemOpen
  \bibfield  {author} {\bibinfo {author} {\bibfnamefont {J.~D.}\ \bibnamefont
  {Strand}}, \bibinfo {author} {\bibfnamefont {M.}~\bibnamefont {Ware}},
  \bibinfo {author} {\bibfnamefont {F.}~\bibnamefont {Beaudoin}}, \bibinfo
  {author} {\bibfnamefont {T.~A.}\ \bibnamefont {Ohki}}, \bibinfo {author}
  {\bibfnamefont {B.~R.}\ \bibnamefont {Johnson}}, \bibinfo {author}
  {\bibfnamefont {A.}~\bibnamefont {Blais}},\ and\ \bibinfo {author}
  {\bibfnamefont {B.~L.}\ \bibnamefont {Plourde}},\ }\bibfield  {title}
  {\bibinfo {title} {{First-order sideband transitions with flux-driven
  asymmetric transmon qubits}},\ }\href
  {https://doi.org/10.1103/PhysRevB.87.220505} {\bibfield  {journal} {\bibinfo
  {journal} {Physical Review B - Condensed Matter and Materials Physics}\
  }\textbf {\bibinfo {volume} {87}},\ \bibinfo {pages} {220505} (\bibinfo
  {year} {2013})}\BibitemShut {NoStop}%
\bibitem [{\citenamefont {Roth}\ \emph {et~al.}(2017)\citenamefont {Roth},
  \citenamefont {Ganzhorn}, \citenamefont {Moll}, \citenamefont {Filipp},
  \citenamefont {Salis},\ and\ \citenamefont {Schmidt}}]{Roth2017}%
  \BibitemOpen
  \bibfield  {author} {\bibinfo {author} {\bibfnamefont {M.}~\bibnamefont
  {Roth}}, \bibinfo {author} {\bibfnamefont {M.}~\bibnamefont {Ganzhorn}},
  \bibinfo {author} {\bibfnamefont {N.}~\bibnamefont {Moll}}, \bibinfo {author}
  {\bibfnamefont {S.}~\bibnamefont {Filipp}}, \bibinfo {author} {\bibfnamefont
  {G.}~\bibnamefont {Salis}},\ and\ \bibinfo {author} {\bibfnamefont
  {S.}~\bibnamefont {Schmidt}},\ }\bibfield  {title} {\bibinfo {title}
  {{Analysis of a parametrically driven exchange-type gate and a two-photon
  excitation gate between superconducting qubits}},\ }\href
  {https://doi.org/10.1103/PhysRevA.96.062323} {\bibfield  {journal} {\bibinfo
  {journal} {Physical Review A}\ }\textbf {\bibinfo {volume} {96}},\ \bibinfo
  {pages} {062323} (\bibinfo {year} {2017})}\BibitemShut {NoStop}%
\bibitem [{\citenamefont {Wallraff}\ \emph {et~al.}(2007)\citenamefont
  {Wallraff}, \citenamefont {Schuster}, \citenamefont {Blais}, \citenamefont
  {Gambetta}, \citenamefont {Schreier}, \citenamefont {Frunzio}, \citenamefont
  {Devoret}, \citenamefont {Girvin},\ and\ \citenamefont
  {Schoelkopf}}]{Wallraff2007}%
  \BibitemOpen
  \bibfield  {author} {\bibinfo {author} {\bibfnamefont {A.}~\bibnamefont
  {Wallraff}}, \bibinfo {author} {\bibfnamefont {D.~I.}\ \bibnamefont
  {Schuster}}, \bibinfo {author} {\bibfnamefont {A.}~\bibnamefont {Blais}},
  \bibinfo {author} {\bibfnamefont {J.~M.}\ \bibnamefont {Gambetta}}, \bibinfo
  {author} {\bibfnamefont {J.}~\bibnamefont {Schreier}}, \bibinfo {author}
  {\bibfnamefont {L.}~\bibnamefont {Frunzio}}, \bibinfo {author} {\bibfnamefont
  {M.~H.}\ \bibnamefont {Devoret}}, \bibinfo {author} {\bibfnamefont {S.~M.}\
  \bibnamefont {Girvin}},\ and\ \bibinfo {author} {\bibfnamefont {R.~J.}\
  \bibnamefont {Schoelkopf}},\ }\bibfield  {title} {\bibinfo {title} {{Sideband
  transitions and two-tone spectroscopy of a superconducting qubit strongly
  coupled to an on-chip cavity}},\ }\href
  {https://doi.org/10.1103/PhysRevLett.99.050501} {\bibfield  {journal}
  {\bibinfo  {journal} {Physical Review Letters}\ }\textbf {\bibinfo {volume}
  {99}},\ \bibinfo {pages} {050501} (\bibinfo {year} {2007})}\BibitemShut
  {NoStop}%
\bibitem [{\citenamefont {Leek}\ \emph {et~al.}(2009)\citenamefont {Leek},
  \citenamefont {Filipp}, \citenamefont {Maurer}, \citenamefont {Baur},
  \citenamefont {Bianchetti}, \citenamefont {Fink}, \citenamefont
  {G{\"{o}}ppl}, \citenamefont {Steffen},\ and\ \citenamefont
  {Wallraff}}]{Leek2009}%
  \BibitemOpen
  \bibfield  {author} {\bibinfo {author} {\bibfnamefont {P.~J.}\ \bibnamefont
  {Leek}}, \bibinfo {author} {\bibfnamefont {S.}~\bibnamefont {Filipp}},
  \bibinfo {author} {\bibfnamefont {P.}~\bibnamefont {Maurer}}, \bibinfo
  {author} {\bibfnamefont {M.}~\bibnamefont {Baur}}, \bibinfo {author}
  {\bibfnamefont {R.}~\bibnamefont {Bianchetti}}, \bibinfo {author}
  {\bibfnamefont {J.~M.}\ \bibnamefont {Fink}}, \bibinfo {author}
  {\bibfnamefont {M.}~\bibnamefont {G{\"{o}}ppl}}, \bibinfo {author}
  {\bibfnamefont {L.}~\bibnamefont {Steffen}},\ and\ \bibinfo {author}
  {\bibfnamefont {A.}~\bibnamefont {Wallraff}},\ }\bibfield  {title} {\bibinfo
  {title} {{Using sideband transitions for two-qubit operations in
  superconducting circuits}},\ }\href
  {https://doi.org/10.1103/PhysRevB.79.180511} {\bibfield  {journal} {\bibinfo
  {journal} {Physical Review B - Condensed Matter and Materials Physics}\
  }\textbf {\bibinfo {volume} {79}},\ \bibinfo {pages} {180511} (\bibinfo
  {year} {2009})}\BibitemShut {NoStop}%
\bibitem [{\citenamefont {Novikov}\ \emph {et~al.}(2016)\citenamefont
  {Novikov}, \citenamefont {Sweeney}, \citenamefont {Robinson}, \citenamefont
  {Premaratne}, \citenamefont {Suri}, \citenamefont {Wellstood},\ and\
  \citenamefont {Palmer}}]{Novikov2016}%
  \BibitemOpen
  \bibfield  {author} {\bibinfo {author} {\bibfnamefont {S.}~\bibnamefont
  {Novikov}}, \bibinfo {author} {\bibfnamefont {T.}~\bibnamefont {Sweeney}},
  \bibinfo {author} {\bibfnamefont {J.~E.}\ \bibnamefont {Robinson}}, \bibinfo
  {author} {\bibfnamefont {S.~P.}\ \bibnamefont {Premaratne}}, \bibinfo
  {author} {\bibfnamefont {B.}~\bibnamefont {Suri}}, \bibinfo {author}
  {\bibfnamefont {F.~C.}\ \bibnamefont {Wellstood}},\ and\ \bibinfo {author}
  {\bibfnamefont {B.~S.}\ \bibnamefont {Palmer}},\ }\bibfield  {title}
  {\bibinfo {title} {{Raman coherence in a circuit quantum electrodynamics
  lambda system}},\ }\href {https://doi.org/10.1038/nphys3537} {\bibfield
  {journal} {\bibinfo  {journal} {Nature Physics}\ }\textbf {\bibinfo {volume}
  {12}},\ \bibinfo {pages} {75} (\bibinfo {year} {2016})}\BibitemShut {NoStop}%
\bibitem [{\citenamefont {Lu}\ \emph {et~al.}(2017)\citenamefont {Lu},
  \citenamefont {Chakram}, \citenamefont {Leung}, \citenamefont {Earnest},
  \citenamefont {Naik}, \citenamefont {Huang}, \citenamefont {Groszkowski},
  \citenamefont {Kapit}, \citenamefont {Koch},\ and\ \citenamefont
  {Schuster}}]{Lu2017}%
  \BibitemOpen
  \bibfield  {author} {\bibinfo {author} {\bibfnamefont {Y.}~\bibnamefont
  {Lu}}, \bibinfo {author} {\bibfnamefont {S.}~\bibnamefont {Chakram}},
  \bibinfo {author} {\bibfnamefont {N.}~\bibnamefont {Leung}}, \bibinfo
  {author} {\bibfnamefont {N.}~\bibnamefont {Earnest}}, \bibinfo {author}
  {\bibfnamefont {R.~K.}\ \bibnamefont {Naik}}, \bibinfo {author}
  {\bibfnamefont {Z.}~\bibnamefont {Huang}}, \bibinfo {author} {\bibfnamefont
  {P.}~\bibnamefont {Groszkowski}}, \bibinfo {author} {\bibfnamefont
  {E.}~\bibnamefont {Kapit}}, \bibinfo {author} {\bibfnamefont
  {J.}~\bibnamefont {Koch}},\ and\ \bibinfo {author} {\bibfnamefont {D.~I.}\
  \bibnamefont {Schuster}},\ }\bibfield  {title} {\bibinfo {title} {{Universal
  Stabilization of a Parametrically Coupled Qubit}},\ }\href@noop {} {\bibfield
   {journal} {\bibinfo  {journal} {Physical Review Letters}\ }\textbf {\bibinfo
  {volume} {119}} (\bibinfo {year} {2017})}\BibitemShut {NoStop}%
\bibitem [{\citenamefont {Huang}\ \emph {et~al.}(2018)\citenamefont {Huang},
  \citenamefont {Lu}, \citenamefont {Kapit}, \citenamefont {Schuster},\ and\
  \citenamefont {Koch}}]{Huang2018}%
  \BibitemOpen
  \bibfield  {author} {\bibinfo {author} {\bibfnamefont {Z.}~\bibnamefont
  {Huang}}, \bibinfo {author} {\bibfnamefont {Y.}~\bibnamefont {Lu}}, \bibinfo
  {author} {\bibfnamefont {E.}~\bibnamefont {Kapit}}, \bibinfo {author}
  {\bibfnamefont {D.~I.}\ \bibnamefont {Schuster}},\ and\ \bibinfo {author}
  {\bibfnamefont {J.}~\bibnamefont {Koch}},\ }\bibfield  {title} {\bibinfo
  {title} {{Universal stabilization of single-qubit states using a tunable
  coupler}},\ }\href {https://doi.org/10.1103/PhysRevA.97.062345} {\bibfield
  {journal} {\bibinfo  {journal} {Physical Review A}\ }\textbf {\bibinfo
  {volume} {97}},\ \bibinfo {pages} {1} (\bibinfo {year} {2018})}\BibitemShut
  {NoStop}%
\bibitem [{\citenamefont {Magnard}\ \emph {et~al.}(2018)\citenamefont
  {Magnard}, \citenamefont {Kurpiers}, \citenamefont {Royer}, \citenamefont
  {Walter}, \citenamefont {Besse}, \citenamefont {Gasparinetti}, \citenamefont
  {Pechal}, \citenamefont {Heinsoo}, \citenamefont {Storz}, \citenamefont
  {Blais},\ and\ \citenamefont {Wallraff}}]{Magnard2018}%
  \BibitemOpen
  \bibfield  {author} {\bibinfo {author} {\bibfnamefont {P.}~\bibnamefont
  {Magnard}}, \bibinfo {author} {\bibfnamefont {P.}~\bibnamefont {Kurpiers}},
  \bibinfo {author} {\bibfnamefont {B.}~\bibnamefont {Royer}}, \bibinfo
  {author} {\bibfnamefont {T.}~\bibnamefont {Walter}}, \bibinfo {author}
  {\bibfnamefont {J.~C.}\ \bibnamefont {Besse}}, \bibinfo {author}
  {\bibfnamefont {S.}~\bibnamefont {Gasparinetti}}, \bibinfo {author}
  {\bibfnamefont {M.}~\bibnamefont {Pechal}}, \bibinfo {author} {\bibfnamefont
  {J.}~\bibnamefont {Heinsoo}}, \bibinfo {author} {\bibfnamefont
  {S.}~\bibnamefont {Storz}}, \bibinfo {author} {\bibfnamefont
  {A.}~\bibnamefont {Blais}},\ and\ \bibinfo {author} {\bibfnamefont
  {A.}~\bibnamefont {Wallraff}},\ }\bibfield  {title} {\bibinfo {title} {{Fast
  and Unconditional All-Microwave Reset of a Superconducting Qubit}},\
  }\href@noop {} {\bibfield  {journal} {\bibinfo  {journal} {Physical Review
  Letters}\ }\textbf {\bibinfo {volume} {121}} (\bibinfo {year}
  {2018})}\BibitemShut {NoStop}%
\bibitem [{\citenamefont {Reed}\ \emph {et~al.}(2010)\citenamefont {Reed},
  \citenamefont {Johnson}, \citenamefont {Houck}, \citenamefont {Dicarlo},
  \citenamefont {Chow}, \citenamefont {Schuster}, \citenamefont {Frunzio},\
  and\ \citenamefont {Schoelkopf}}]{Reed2010}%
  \BibitemOpen
  \bibfield  {author} {\bibinfo {author} {\bibfnamefont {M.~D.}\ \bibnamefont
  {Reed}}, \bibinfo {author} {\bibfnamefont {B.~R.}\ \bibnamefont {Johnson}},
  \bibinfo {author} {\bibfnamefont {A.~A.}\ \bibnamefont {Houck}}, \bibinfo
  {author} {\bibfnamefont {L.}~\bibnamefont {Dicarlo}}, \bibinfo {author}
  {\bibfnamefont {J.~M.}\ \bibnamefont {Chow}}, \bibinfo {author}
  {\bibfnamefont {D.~I.}\ \bibnamefont {Schuster}}, \bibinfo {author}
  {\bibfnamefont {L.}~\bibnamefont {Frunzio}},\ and\ \bibinfo {author}
  {\bibfnamefont {R.~J.}\ \bibnamefont {Schoelkopf}},\ }\bibfield  {title}
  {\bibinfo {title} {{Fast reset and suppressing spontaneous emission of a
  superconducting qubit}},\ }\href {https://doi.org/10.1063/1.3435463}
  {\bibfield  {journal} {\bibinfo  {journal} {Applied Physics Letters}\
  }\textbf {\bibinfo {volume} {96}},\ \bibinfo {pages} {203110} (\bibinfo
  {year} {2010})}\BibitemShut {NoStop}%
\bibitem [{\citenamefont {Valenzuela}\ \emph {et~al.}(2006)\citenamefont
  {Valenzuela}, \citenamefont {Oliver}, \citenamefont {Berns}, \citenamefont
  {Berggren}, \citenamefont {Levitov},\ and\ \citenamefont
  {Orlando}}]{Valenzuela2006}%
  \BibitemOpen
  \bibfield  {author} {\bibinfo {author} {\bibfnamefont {S.~O.}\ \bibnamefont
  {Valenzuela}}, \bibinfo {author} {\bibfnamefont {W.~D.}\ \bibnamefont
  {Oliver}}, \bibinfo {author} {\bibfnamefont {D.~M.}\ \bibnamefont {Berns}},
  \bibinfo {author} {\bibfnamefont {K.~K.}\ \bibnamefont {Berggren}}, \bibinfo
  {author} {\bibfnamefont {L.~S.}\ \bibnamefont {Levitov}},\ and\ \bibinfo
  {author} {\bibfnamefont {T.~P.}\ \bibnamefont {Orlando}},\ }\bibfield
  {title} {\bibinfo {title} {{Microwave-induced cooling of a superconducting
  qubit}},\ }\href {https://doi.org/10.1126/science.1134008} {\bibfield
  {journal} {\bibinfo  {journal} {Science}\ }\textbf {\bibinfo {volume}
  {314}},\ \bibinfo {pages} {1589} (\bibinfo {year} {2006})}\BibitemShut
  {NoStop}%
\bibitem [{\citenamefont {Geerlings}\ \emph {et~al.}(2013)\citenamefont
  {Geerlings}, \citenamefont {Leghtas}, \citenamefont {Pop}, \citenamefont
  {Shankar}, \citenamefont {Frunzio}, \citenamefont {Schoelkopf}, \citenamefont
  {Mirrahimi},\ and\ \citenamefont {Devoret}}]{Geerlings2013}%
  \BibitemOpen
  \bibfield  {author} {\bibinfo {author} {\bibfnamefont {K.}~\bibnamefont
  {Geerlings}}, \bibinfo {author} {\bibfnamefont {Z.}~\bibnamefont {Leghtas}},
  \bibinfo {author} {\bibfnamefont {I.~M.}\ \bibnamefont {Pop}}, \bibinfo
  {author} {\bibfnamefont {S.}~\bibnamefont {Shankar}}, \bibinfo {author}
  {\bibfnamefont {L.}~\bibnamefont {Frunzio}}, \bibinfo {author} {\bibfnamefont
  {R.~J.}\ \bibnamefont {Schoelkopf}}, \bibinfo {author} {\bibfnamefont
  {M.}~\bibnamefont {Mirrahimi}},\ and\ \bibinfo {author} {\bibfnamefont
  {M.~H.}\ \bibnamefont {Devoret}},\ }\bibfield  {title} {\bibinfo {title}
  {{Demonstrating a driven reset protocol for a superconducting qubit}},\
  }\href {https://doi.org/10.1103/PhysRevLett.110.120501} {\bibfield  {journal}
  {\bibinfo  {journal} {Physical Review Letters}\ }\textbf {\bibinfo {volume}
  {110}},\ \bibinfo {pages} {120501} (\bibinfo {year} {2013})}\BibitemShut
  {NoStop}%
\bibitem [{\citenamefont {Ma}\ \emph {et~al.}(2017)\citenamefont {Ma},
  \citenamefont {Owens}, \citenamefont {Houck}, \citenamefont {Schuster},\ and\
  \citenamefont {Simon}}]{Ma2017}%
  \BibitemOpen
  \bibfield  {author} {\bibinfo {author} {\bibfnamefont {R.}~\bibnamefont
  {Ma}}, \bibinfo {author} {\bibfnamefont {C.}~\bibnamefont {Owens}}, \bibinfo
  {author} {\bibfnamefont {A.}~\bibnamefont {Houck}}, \bibinfo {author}
  {\bibfnamefont {D.~I.}\ \bibnamefont {Schuster}},\ and\ \bibinfo {author}
  {\bibfnamefont {J.}~\bibnamefont {Simon}},\ }\bibfield  {title} {\bibinfo
  {title} {{Autonomous stabilizer for incompressible photon fluids and
  solids}},\ }\href@noop {} {\bibfield  {journal} {\bibinfo  {journal}
  {Physical Review A}\ }\textbf {\bibinfo {volume} {95}} (\bibinfo {year}
  {2017})}\BibitemShut {NoStop}%
\bibitem [{\citenamefont {Khaneja}\ \emph {et~al.}(2005)\citenamefont
  {Khaneja}, \citenamefont {Reiss}, \citenamefont {Kehlet}, \citenamefont
  {Schulte-Herbr{\"{u}}ggen},\ and\ \citenamefont {Glaser}}]{Khaneja2005}%
  \BibitemOpen
  \bibfield  {author} {\bibinfo {author} {\bibfnamefont {N.}~\bibnamefont
  {Khaneja}}, \bibinfo {author} {\bibfnamefont {T.}~\bibnamefont {Reiss}},
  \bibinfo {author} {\bibfnamefont {C.}~\bibnamefont {Kehlet}}, \bibinfo
  {author} {\bibfnamefont {T.}~\bibnamefont {Schulte-Herbr{\"{u}}ggen}},\ and\
  \bibinfo {author} {\bibfnamefont {S.~J.}\ \bibnamefont {Glaser}},\ }\bibfield
   {title} {\bibinfo {title} {{Optimal control of coupled spin dynamics: Design
  of NMR pulse sequences by gradient ascent algorithms}},\ }\href
  {https://doi.org/10.1016/j.jmr.2004.11.004} {\bibfield  {journal} {\bibinfo
  {journal} {Journal of Magnetic Resonance}\ }\textbf {\bibinfo {volume}
  {172}},\ \bibinfo {pages} {296} (\bibinfo {year} {2005})}\BibitemShut
  {NoStop}%
\bibitem [{\citenamefont {Motzoi}\ \emph {et~al.}(2009)\citenamefont {Motzoi},
  \citenamefont {Gambetta}, \citenamefont {Rebentrost},\ and\ \citenamefont
  {Wilhelm}}]{Motzoi2009}%
  \BibitemOpen
  \bibfield  {author} {\bibinfo {author} {\bibfnamefont {F.}~\bibnamefont
  {Motzoi}}, \bibinfo {author} {\bibfnamefont {J.~M.}\ \bibnamefont
  {Gambetta}}, \bibinfo {author} {\bibfnamefont {P.}~\bibnamefont
  {Rebentrost}},\ and\ \bibinfo {author} {\bibfnamefont {F.~K.}\ \bibnamefont
  {Wilhelm}},\ }\bibfield  {title} {\bibinfo {title} {{Simple Pulses for
  Elimination of Leakage in Weakly Nonlinear Qubits}},\ }\href@noop {}
  {\bibfield  {journal} {\bibinfo  {journal} {Physical Review Letters}\
  }\textbf {\bibinfo {volume} {103}} (\bibinfo {year} {2009})}\BibitemShut
  {NoStop}%
\bibitem [{\citenamefont {M{\"{o}}tt{\"{o}}nen}\ \emph
  {et~al.}(2006)\citenamefont {M{\"{o}}tt{\"{o}}nen}, \citenamefont {{De
  Sousa}}, \citenamefont {Zhang},\ and\ \citenamefont {Whaley}}]{Mottonen2006}%
  \BibitemOpen
  \bibfield  {author} {\bibinfo {author} {\bibfnamefont {M.}~\bibnamefont
  {M{\"{o}}tt{\"{o}}nen}}, \bibinfo {author} {\bibfnamefont {R.}~\bibnamefont
  {{De Sousa}}}, \bibinfo {author} {\bibfnamefont {J.}~\bibnamefont {Zhang}},\
  and\ \bibinfo {author} {\bibfnamefont {K.~B.}\ \bibnamefont {Whaley}},\
  }\bibfield  {title} {\bibinfo {title} {{High-fidelity one-qubit operations
  under random telegraph noise}},\ }\href@noop {} {\bibfield  {journal}
  {\bibinfo  {journal} {Physical Review A - Atomic, Molecular, and Optical
  Physics}\ }\textbf {\bibinfo {volume} {73}} (\bibinfo {year}
  {2006})}\BibitemShut {NoStop}%
\bibitem [{\citenamefont {Steffen}\ \emph {et~al.}(2003)\citenamefont
  {Steffen}, \citenamefont {Martinis},\ and\ \citenamefont {Chuang}}]{Steffen}%
  \BibitemOpen
  \bibfield  {author} {\bibinfo {author} {\bibfnamefont {M.}~\bibnamefont
  {Steffen}}, \bibinfo {author} {\bibfnamefont {J.~M.}\ \bibnamefont
  {Martinis}},\ and\ \bibinfo {author} {\bibfnamefont {I.~L.}\ \bibnamefont
  {Chuang}},\ }\bibfield  {title} {\bibinfo {title} {{Accurate control of
  Josephson phase qubits}},\ }\href@noop {} {\bibfield  {journal} {\bibinfo
  {journal} {Physical Review B - Condensed Matter and Materials Physics}\
  }\textbf {\bibinfo {volume} {68}} (\bibinfo {year} {2003})}\BibitemShut
  {NoStop}%
\bibitem [{\citenamefont {Safaei}\ \emph {et~al.}(2009)\citenamefont {Safaei},
  \citenamefont {Montangero}, \citenamefont {Taddei},\ and\ \citenamefont
  {Fazio}}]{Safaei2009}%
  \BibitemOpen
  \bibfield  {author} {\bibinfo {author} {\bibfnamefont {S.}~\bibnamefont
  {Safaei}}, \bibinfo {author} {\bibfnamefont {S.}~\bibnamefont {Montangero}},
  \bibinfo {author} {\bibfnamefont {F.}~\bibnamefont {Taddei}},\ and\ \bibinfo
  {author} {\bibfnamefont {R.}~\bibnamefont {Fazio}},\ }\bibfield  {title}
  {\bibinfo {title} {{Optimized single-qubit gates for Josephson phase
  qubits}},\ }\href {https://doi.org/10.1103/PhysRevB.79.064524} {\bibfield
  {journal} {\bibinfo  {journal} {Physical Review B - Condensed Matter and
  Materials Physics}\ }\textbf {\bibinfo {volume} {79}},\ \bibinfo {pages}
  {064524} (\bibinfo {year} {2009})}\BibitemShut {NoStop}%
\bibitem [{\citenamefont {Pravia}\ \emph {et~al.}(2003)\citenamefont {Pravia},
  \citenamefont {Boulant}, \citenamefont {Emerson}, \citenamefont {Farid},
  \citenamefont {Fortunato}, \citenamefont {Havel}, \citenamefont {Martinez},\
  and\ \citenamefont {Cory}}]{Pravia2003}%
  \BibitemOpen
  \bibfield  {author} {\bibinfo {author} {\bibfnamefont {M.~A.}\ \bibnamefont
  {Pravia}}, \bibinfo {author} {\bibfnamefont {N.}~\bibnamefont {Boulant}},
  \bibinfo {author} {\bibfnamefont {J.}~\bibnamefont {Emerson}}, \bibinfo
  {author} {\bibfnamefont {A.}~\bibnamefont {Farid}}, \bibinfo {author}
  {\bibfnamefont {E.~M.}\ \bibnamefont {Fortunato}}, \bibinfo {author}
  {\bibfnamefont {T.~F.}\ \bibnamefont {Havel}}, \bibinfo {author}
  {\bibfnamefont {R.}~\bibnamefont {Martinez}},\ and\ \bibinfo {author}
  {\bibfnamefont {D.~G.}\ \bibnamefont {Cory}},\ }\bibfield  {title} {\bibinfo
  {title} {{Incoherent noise and quantum information processing}},\ }\href
  {https://doi.org/10.1063/1.1619132} {\bibfield  {journal} {\bibinfo
  {journal} {Quantum Computation and Quantum Information American Journal of
  Physics}\ }\textbf {\bibinfo {volume} {119}},\ \bibinfo {pages} {558}
  (\bibinfo {year} {2003})}\BibitemShut {NoStop}%
\bibitem [{\citenamefont {Reed}\ \emph {et~al.}(2012)\citenamefont {Reed},
  \citenamefont {DiCarlo}, \citenamefont {Nigg}, \citenamefont {Sun},
  \citenamefont {Frunzio}, \citenamefont {Girvin},\ and\ \citenamefont
  {Schoelkopf}}]{Reed2012}%
  \BibitemOpen
  \bibfield  {author} {\bibinfo {author} {\bibfnamefont {M.~D.}\ \bibnamefont
  {Reed}}, \bibinfo {author} {\bibfnamefont {L.}~\bibnamefont {DiCarlo}},
  \bibinfo {author} {\bibfnamefont {S.~E.}\ \bibnamefont {Nigg}}, \bibinfo
  {author} {\bibfnamefont {L.}~\bibnamefont {Sun}}, \bibinfo {author}
  {\bibfnamefont {L.}~\bibnamefont {Frunzio}}, \bibinfo {author} {\bibfnamefont
  {S.~M.}\ \bibnamefont {Girvin}},\ and\ \bibinfo {author} {\bibfnamefont
  {R.~J.}\ \bibnamefont {Schoelkopf}},\ }\bibfield  {title} {\bibinfo {title}
  {{Realization of three-qubit quantum error correction with superconducting
  circuits}},\ }\href {https://doi.org/10.1038/nature10786} {\bibfield
  {journal} {\bibinfo  {journal} {Nature}\ }\textbf {\bibinfo {volume} {482}},\
  \bibinfo {pages} {382} (\bibinfo {year} {2012})}\BibitemShut {NoStop}%
\bibitem [{\citenamefont {Cohen}\ and\ \citenamefont
  {Mirrahimi}(2014)}]{Cohen2014}%
  \BibitemOpen
  \bibfield  {author} {\bibinfo {author} {\bibfnamefont {J.}~\bibnamefont
  {Cohen}}\ and\ \bibinfo {author} {\bibfnamefont {M.}~\bibnamefont
  {Mirrahimi}},\ }\bibfield  {title} {\bibinfo {title} {{Dissipation-induced
  continuous quantum error correction for superconducting circuits}},\ }\href
  {https://doi.org/10.1103/PhysRevA.90.062344} {\bibfield  {journal} {\bibinfo
  {journal} {Physical Review A}\ }\textbf {\bibinfo {volume} {90}},\ \bibinfo
  {pages} {062344} (\bibinfo {year} {2014})}\BibitemShut {NoStop}%
\bibitem [{\citenamefont {Kapit}\ \emph {et~al.}(2014)\citenamefont {Kapit},
  \citenamefont {Hafezi},\ and\ \citenamefont {Simon}}]{Kapit2014}%
  \BibitemOpen
  \bibfield  {author} {\bibinfo {author} {\bibfnamefont {E.}~\bibnamefont
  {Kapit}}, \bibinfo {author} {\bibfnamefont {M.}~\bibnamefont {Hafezi}},\ and\
  \bibinfo {author} {\bibfnamefont {S.~H.}\ \bibnamefont {Simon}},\ }\bibfield
  {title} {\bibinfo {title} {{Induced Self-Stabilization in Fractional Quantum
  Hall States of Light}},\ }\href {https://doi.org/10.1103/PhysRevX.4.031039}
  {\bibfield  {journal} {\bibinfo  {journal} {Physical Review X}\ }\textbf
  {\bibinfo {volume} {4}},\ \bibinfo {pages} {031039} (\bibinfo {year}
  {2014})}\BibitemShut {NoStop}%
\bibitem [{\citenamefont {Kapit}(2015)}]{Kapit2015}%
  \BibitemOpen
  \bibfield  {author} {\bibinfo {author} {\bibfnamefont {E.}~\bibnamefont
  {Kapit}},\ }\bibfield  {title} {\bibinfo {title} {{Universal two-qubit
  interactions, measurement, and cooling for quantum simulation and
  computing}},\ }\href@noop {} {\bibfield  {journal} {\bibinfo  {journal}
  {Physical Review A - Atomic, Molecular, and Optical Physics}\ }\textbf
  {\bibinfo {volume} {92}},\ \bibinfo {pages} {012302} (\bibinfo {year}
  {2015})}\BibitemShut {NoStop}%
\bibitem [{\citenamefont {Kapit}(2018)}]{Kapit2018}%
  \BibitemOpen
  \bibfield  {author} {\bibinfo {author} {\bibfnamefont {E.}~\bibnamefont
  {Kapit}},\ }\bibfield  {title} {\bibinfo {title} {{Error-Transparent Quantum
  Gates for Small Logical Qubit Architectures}},\ }\href
  {https://doi.org/10.1103/PhysRevLett.120.050503} {\bibfield  {journal}
  {\bibinfo  {journal} {Physical Review Letters}\ }\textbf {\bibinfo {volume}
  {120}},\ \bibinfo {pages} {50503} (\bibinfo {year} {2018})}\BibitemShut
  {NoStop}%
\end{thebibliography}%

\end{document}